\documentclass[a4paper,fleqn,usenatbib]{mnras}
\usepackage[T1]{fontenc}
\usepackage{ae,aecompl}
\usepackage{graphicx}	
\usepackage{amsmath}	
\usepackage{amssymb}	
\usepackage{color,ulem}
\newcommand{\Vrel}{V_\mathrm{rel}}    
\newcommand{\Vfrag}{V_\mathrm{frag}}  
\newcommand{\St}{\mathrm{St}}         
\title{Self-induced dust traps: overcoming planet formation barriers}

\author[J.-F. Gonzalez et al.]{
J.-F. Gonzalez,$^{1}$\thanks{E-mail: jean-francois.gonzalez@ens-lyon.fr}
G. Laibe,$^{1,2}$\thanks{E-mail: guillaume.laibe@ens-lyon.fr}
and S. T. Maddison,$^{3}$
\\
$^{1}$Univ Lyon, Univ Lyon1, Ens de Lyon, CNRS, Centre de Recherche Astrophysique de Lyon UMR5574, F-69230, Saint-Genis-Laval, France\\
$^{2}$School of Physics and Astronomy, University of Saint Andrews,
North Haugh, St Andrews, Fife KY16 9SS, United Kingdom\\
$^{3}$Centre for Astrophysics and Supercomputing, Swinburne University of Technology, PO Box 218, Hawthorn, VIC 3122, Australia
}

\date{Accepted 2017 January 4. Received 2016 December 8; in original form 2016 July 5.}

\pubyear{2017}

\begin{document}
\label{firstpage}
\pagerange{\pageref{firstpage}--\pageref{lastpage}}
\maketitle

\begin{abstract}
Planet formation is thought to occur in discs around young stars by the aggregation of small dust grains into much larger objects. The growth from grains to pebbles and from planetesimals to planets is now fairly well understood. The intermediate stage has however been found to be hindered by the radial-drift and fragmentation barriers. We identify a powerful mechanism in which dust overcomes both barriers. Its key ingredients are i) backreaction from the dust onto the gas, ii) grain growth and fragmentation, and iii) large-scale gradients. The pile-up of growing and fragmenting grains modifies the gas structure on large scales and triggers the formation of pressure maxima, in which particles are trapped. We show that these self-induced dust traps are robust: they develop for a wide range of disc structures, fragmentation thresholds and initial dust-to-gas ratios. They are favored locations for pebbles to grow into planetesimals, thus opening new paths towards the formation of planets.
\end{abstract}

\begin{keywords}
Protoplanetary discs -- Hydrodynamics -- Methods: numerical
\end{keywords}



\section{Introduction}
\label{sec:Introduction}

In the core-accretion paradigm, planets are thought to originate from solid cores which accrete the surrounding material of their protoplanetary discs \citep{Lissauer1993,Pollack1996,Alibert2005,Mordasini2012}. This requires primordial dust grains to concentrate and grow to form the building blocks of planets \citep{Dominik2007}. However, there are two important barriers that need to be overcome for this to happen. Grains feel the aerodynamic drag of the gas in the disc, causing the grains to settle to the midplane and drift inwards as they lose angular momentum. The efficiency of this relative motion of dust with respect to gas is controlled by the Stokes number St, i.e. the ratio of the drag stopping time to the orbital time, which depends on grain size and local gas conditions. Small ($\St\ll1$, strong gas-dust coupling) and large ($\St\gg1$, weak coupling) grains drift slowly, whereas grains with $\mathrm{St}\sim1$ (i.e. millimetre-sized at a few tens of au in observed discs, see e.g.\ \citealt{Laibe2012,Dipierro2015}) experience the fastest radial drift, which can lead to their accretion onto the star before growing into planet embryos \citep{Adachi1976,Weidenschilling1977}. Efficient grain growth is needed for dust particles to overcome this so-called `radial-drift barrier' \citep{LGM2014}. Larger decimetre-sized grains, on the other hand, which have higher relative velocities \citep{WC1993}, shatter rather than stick upon collision, leading to the `fragmentation barrier' \citep{Blum2008}. This raises the question: how, in spite of these barriers, do planets manage to form?

Dust grains drift towards the maximum of the gas pressure \citep{Weidenschilling1977}, which for a smooth disc is located at the inner disc edge, resulting in their radial motion towards the central star. Any local gas pressure maximum in the disc will attract and trap grains both in the vicinity and those drifting inwards from the outer regions, forming a so-called dust trap \citep{Whipple1972}. Such a trap suppresses any further radial drift of dust grains. The confinement of solids at this location increases their density, which accelerates grain growth (see Section~\ref{sec:Methods}), and reduces their relative velocities, which in turn prevents grain fragmentation. Dust traps are therefore an efficient way to overcome both the radial-drift and fragmentation barriers. Several types of dust traps have been proposed, e.g.\ vortices \citep{Barge1995,Regaly2012,Meheut2012,Lyra2012,Zhu2014} or planet gap edges \citep{Paardekooper2004,Paardekooper2006,Rice2006,Fouchet2007,Fouchet2010,Ayliffe2012,Pinilla2012,Zhu2012,Zhu2014}. They all require special conditions in the gas pressure profile or an external factor such as the presence of an existing planet.

Due to the high numerical cost, important processes are often neglected when modelling these dust traps. For example, there exist detailed hydrodynamical studies of instabilities in a gas (modelled as a fluid) and dust (represented by Lagrangian particles) disk with self-gravity but in local simulations and without grain growth or fragmentation \citep{Johansen2007} or sophisticated treatments of growth and fragmentation but in 1D simulations of the dust evolution on a static gas background \citep{Birnstiel2010}. More recently, different models of dust evolution have been performed on a 2D static \citep{Drazkowska2013} or 1+1D evolving gas disc \citep{Drazkowska2016}. To date, no numerical study has simultaneously treated all relevant physical ingredients in global 3D simulations of protoplanetary discs. Our two-fluid (gas+dust) Smoothed Particle Hydrodynamics (SPH) code allows global 3D simulations of discs where the coupled evolution of gas and dust is treated self-consistently. We include both the aerodynamic drag of gas on dust and its backreaction, i.e. the drag of dust on gas, as well as the growth and fragmentation of dust grains (see Section~\ref{sec:Methods}). Backreaction is usually ignored because it is negligible when the dust-to-gas ratio $\epsilon$ is low, which is the case most of the time in astrophysical objects. Its effects become important, however, in situations where $\epsilon$ locally approaches unity, like in planet formation. Taking into account these processes simultaneously, in this work we demonstrate that dust traps can develop spontaneously in protoplanetary discs under a variety of conditions.

\begin{figure}
\centering
\resizebox{\hsize}{!}{
\includegraphics{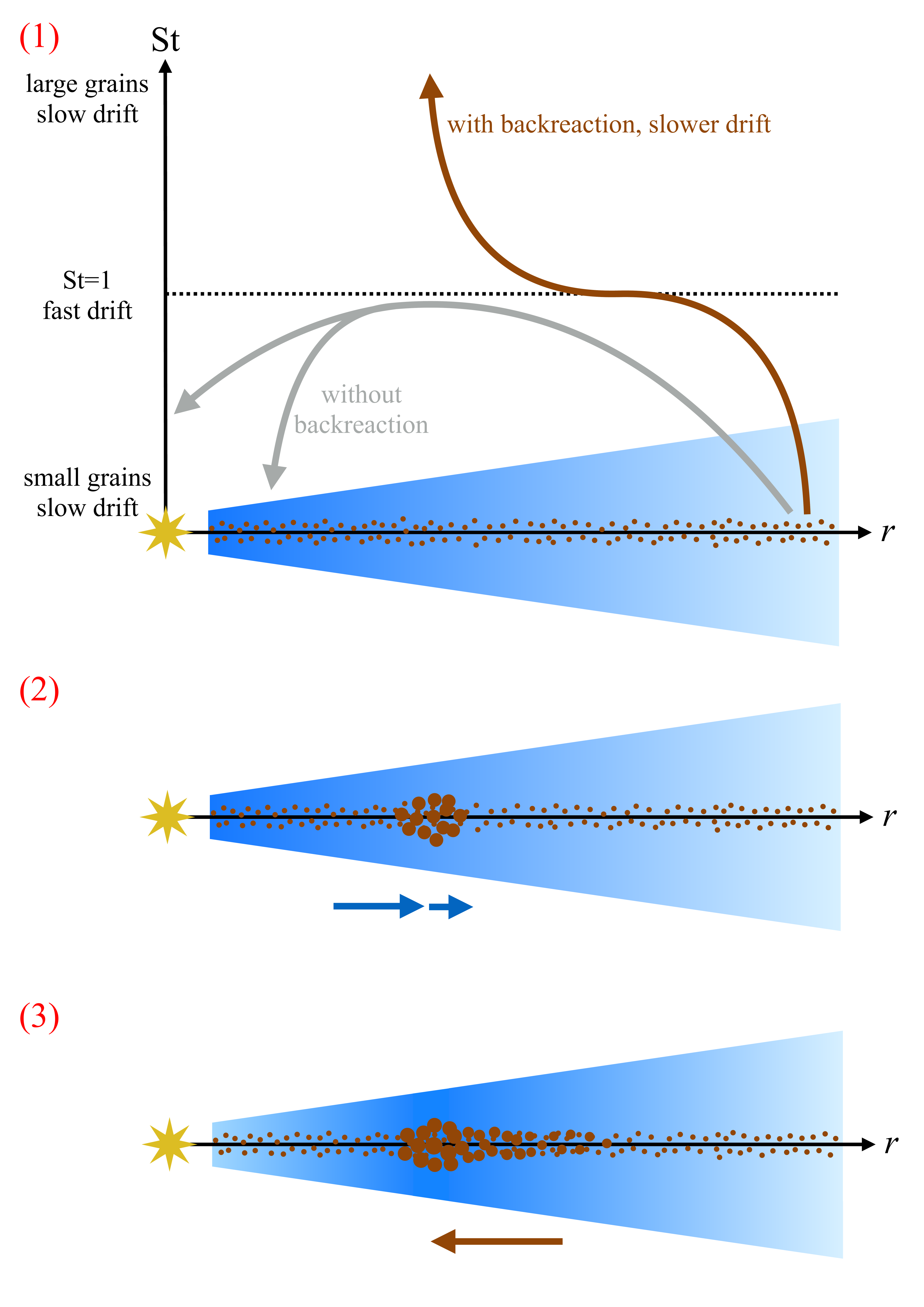}
}
\caption{Self-induced dust trap formation mechanism. (1) Without backreaction, growing grains drift faster and faster as they approach $\St=1$, and either re-fragment in the inner disc or are accreted onto the star (grey). Backreaction slows down the drift, leaving enough time for grains to grow to large sizes, decouple from the gas, and overcome radial drift (brown). (2) Once dust has piled up in a radial concentration, backreaction drags the gas outwards (blue), forming a gas pressure maximum, i.e. a dust trap. (3) The self-induced dust trap stops inward-drifting grains and strengthens the dust concentration.}
\label{Fig:sketch}
\end{figure}

We detail our SPH code and growth/fragmentation model in Section~\ref{sec:Methods} and present the resulting gas and dust distributions in Section~\ref{sec:Results}. We discuss the origin of the self-induced dust traps, whose formation mechanism is summarized in Fig.~\ref{Fig:sketch}, as well as explain how this mechanism differs from the streaming instability in Section~\ref{sec:Discussion}. We summarize our work and conclude in Section~\ref{sec:Conclusion}.

\section{Methods}
\label{sec:Methods}

\subsection{Hydrodynamical simulations}
\label{sec:Hydro}

We use our 3D, two-phase (gas+dust), SPH code to model protoplanetary discs \citep{BF2005}. We adopt a vertically isothermal equation of state which mimics the internal thermal structure of typical discs below a thin surface layer \citep{Woitke2016}. The code does not include self-gravity. This has a negligible effect on our results since the disc-to-star mass ratio is of the order of 1\%\footnote{Only our flat disc model (see Table~\ref{Tab:DiscParam}) is marginally unstable to gravitational instability near its outer rim, which would help planet formation there. However, we are studying a different mechanism and neglecting self-gravity allows to separate the effects.}. Our implementation of gas-dust coupling via aerodynamic drag is that of \citet{Monaghan1997} and conserves the total linear and angular momentum rigorously (the reader is referred to \citealt{Monaghan1997} for general tests and to \citealt{Maddison1998} for additional tests, specific to discs). In particular, we include here the drag of dust on gas, which we call the `backreaction', often neglected in the literature. This term becomes important when properly describing the evolution of the disc midplane, where grains have settled and concentrated to reach dust-to-gas ratios of order unity. In that respect, we obtain larger dust densities in the midplane by treating dust settling dynamically, compared to 2D codes where equilibrium models are used. We adopt the standard SPH artificial viscosity \citep{Monaghan1989} with $\alpha_\mathrm{\scriptscriptstyle AV}=0.1$ and $\beta_\mathrm{\scriptscriptstyle AV}=0.5$, corresponding to a uniform Shakura-Sunyaev \citep{SS1973} $\alpha\sim10^{-2}$, as discussed in \citet{Fouchet2007}. The scheme has been shown to naturally reproduce the expected properties of Prandtl-like turbulence \citep{Arena2013}. In particular, the dust scale height in the stationary regime for grains of constant size reproduces that predicted by \citet{Dubrulle1995}.

\begin{table}
\caption{Parameters of our flat and steep disc models. Values with subscript 0 are given for $r_0=1$~au.}
\label{Tab:DiscParam}
\begin{center}
\begin{tabular}{@{}l@{\quad}c@{\quad}c@{\quad}c@{\quad}c@{\quad}c@{\quad}c@{\quad}c@{\quad}c@{}}
\hline
Model & $p$ & $q$ & $r_\mathrm{in}$ & $r_\mathrm{out}$ & $r_\mathrm{esc}$ & $\Sigma_0$ & $T_0$ & $H_0/R_0$ \\
&&& (au) & (au) & (au) & (kg\,m$^{-2}$) & (K) \\
\hline
flat  & 0 & 1   &  4 & 120 & 160 & 19.67  & 623 & 0.05 \\
steep & 1 & 1/2 & 10 & 300 & 400 & 487.74 & 200 & 0.0283 \\
\hline
\end{tabular}
\end{center}
\end{table}

We simulate discs of mass $M_\mathrm{disc}=0.01\ M_\odot$ orbiting a 1~$M_\odot$ star. Our initial gas disc has a power law surface density $\Sigma_\mathrm{g}=\Sigma_0(r/r_0)^{-p}$ and temperature $T=T_0(r/r_0)^{-q}$. The disc extends from $r_\mathrm{in}$ to $r_\mathrm{out}$ and is allowed to spread. Particles are removed from the simulation if they escape to radii larger than $r_\mathrm{esc}$ or if they drift inside of $r_\mathrm{in}$, where there are assumed to be subsequently accreted by the star. The sound speed varies as $c_\mathrm{s}\propto r^{-\frac{q}{2}}$ and the disc aspect ratio as $H/r\propto r^\frac{1-q}{2}$. We consider two disc models: a flat disc with $p=0$ and $q=1$, commonly used in planet-gap opening studies \citep{Paardekooper2004,Paardekooper2006,Fouchet2007,Fouchet2010,Gonzalez2015a}, and a steep disc with $p=1$ and $q=\frac{1}{2}$, representing an average observed disc from \citet{Williams2014}. The disc radial extension sets the value of $\Sigma_0$. The flat disc has a constant $H/r=0.05$, which sets the temperature scale, whereas the steep disc has $T_0=200$~K at $r_0=1$~au. The relevant model parameters are listed in Table~\ref{Tab:DiscParam}.

We start our simulations with 200,000 gas particles distributed according to the surface density profile with initial Keplerian velocities, and let the gas disc relax for 8 orbits at $r=40$~au for the flat disc and 24 orbits at $r=100$~au for the steep disc. We then overlay the gas particles with an equal number of dust particles, with the same positions and velocities, to reproduce an initially uniform dust-to-gas ratio $\epsilon_0$ and let the disc evolve. The gas particles have equal masses, as do the dust particles, but the two phases differ in mass by a factor $\epsilon_0$. Solids have a typical bulk density $\rho_\mathrm{s}=1000$~kg\,m$^{-3}$, an initial size $s_0=10~\mu$m and are allowed to grow and fragment. The results are insensitive to the value of $s_0$, since small grains initially grow very efficiently \citep{Laibe2008}. Fragmentation of dust grains occurs when dust relative velocities $\Vrel$ reach a threshold $\Vfrag$ which is independent of the grain size. We explore values of $\Vfrag$ of 10, 15, 20 and 25~m\,s$^{-1}$. These values of the fragmentation threshold span the expected range for materials from compact icy grains (10~m\,s$^{-1}$, \citealt{Blum2008}) to porous aggregates (several tens of m\,s$^{-1}$, \citealt{Yamamoto2014}). Since our modeled discs are entirely outside the water ice line, we do not take lower values, typical of compact silicate grains (1~m\,s$^{-1}$, \citealt{Blum2008}), into account. We consider initial dust-to-gas ratios $\epsilon_0$ of 1\%, typical of the interstellar medium and commonly used for protoplanetary discs, for both the flat and steep discs, and higher values of 3 and 5\% in the steep disc model, as indicated by recent observations of a sample of Class~II discs \citep{Williams2014}. The parameters used in the different models are provided in Table~\ref{Tab:Sims}. Simulations are performed up to 200,000~yr for the flat disc and 400,000~yr for the steep disc, to allow the disc form a stationary dust trap.

\begin{table}
\caption{Simulation suite.}
\label{Tab:Sims}
\begin{center}
\begin{tabular}{lccc}
\hline
Model & $\epsilon_0$ & $\Vfrag$ (m\,s$^{-1}$) & backreaction \\
\hline
flat  & 1\% & 10, 15, 20, 25 & yes \\
      & 1\% & 15             & no  \\
\hline
steep & 1\% & 10, 15, 20, 25 & yes \\
      & 1\% & 15             & no  \\
      & 3\% & 10             & yes \\
      & 5\% & 10, 15, 20, 25 & yes \\
\hline
\end{tabular}
\end{center}
\end{table}

\subsection{Drag in the Epstein regime}
\label{sec:Drag}

The gas densities of the discs we model are sufficiently low that the mean free path of gas molecules is larger than the grain sizes \citep{Laibe2012}. Hence, grains are in the Epstein aerodynamic drag regime and the expression of the Stokes number is
\begin{equation}
\St \equiv \frac{\Omega_\mathrm{K}\rho_\mathrm{s}s}{\rho_\mathrm{g}c_\mathrm{s}},
\label{Eq:St}
\end{equation}
where $\Omega_\mathrm{K}$ and $\rho_\mathrm{g}$ denote the Keplerian angular velocity and the gas density, respectively.
Dust grains experiencing the fastest radial drift have $\St = 1$, corresponding to a size
\begin{equation}
s_\mathrm{drift} \equiv \frac{\rho_\mathrm{g}c_\mathrm{s}}{\rho_\mathrm{s}\Omega_\mathrm{K}}.
\label{Eq:s_drift}
\end{equation}
If grains cannot grow fast enough to overcome $\St=1$ before drifting out of the disc, the dust population is drift-limited: $s_\mathrm{drift}$ is the maximum size grains can reach at a given location in the disc.

\subsection{Growth model}
\label{sec:Growth}

Grain growth is implemented as described in \citet{Laibe2008}. The dust size distribution is assumed to be locally monodisperse, i.e. narrowly peaked around a local mean value, and two colliding grains with relative velocities below the fragmentation threshold will stick. The mass $m$ of the resulting larger grain doubles in a mean collision time $\tau_\mathrm{coll}$, i.e. $\mathrm{d}m/\mathrm{d}t=m/\tau_\mathrm{coll}$, which translates to
\begin{equation}
\frac{\mathrm{d}s}{\mathrm{d}t}=\frac{\rho_\mathrm{d}}{\rho_\mathrm{s}}\,\Vrel=\epsilon\,\frac{\rho_\mathrm{g}}{\rho_\mathrm{s}}\,\Vrel,
\label{Eq:dsdt}
\end{equation}
where $\rho_\mathrm{d}$ is the volume density of the dust phase and $\epsilon\equiv\rho_\mathrm{d}/\rho_\mathrm{g}$ the dust-to-gas ratio \citep{SV1997}. The relative velocity between grains $\Vrel$ originates from the gas turbulent motion, which is transferred to dust via aerodynamic drag. We adopt the subgrid model of turbulence developed by \citet{SV1997}, which gives
\begin{equation}
V_\mathrm{rel}=\sqrt{2}\,V_\mathrm{t}\frac{\sqrt{\mathrm{Sc}-1}}{\mathrm{Sc}}.
\label{Eq:Vrel}
\end{equation}
$V_\mathrm{t}\equiv\sqrt{\tilde{\alpha}}c_{\rm s}$ denotes the turbulent velocity, where $\tilde{\alpha}\equiv 2^\frac{1}{2}\,\mathrm{Ro}\,\alpha$ is an effective turbulent diffusivity which depends on Ro, the Rossby number for turbulent motions (taken to be uniform and equal to 3). Sc denotes the Schmidt number of the grains, given by
\begin{equation}
\mathrm{Sc} \equiv (1+\St)\sqrt{1+\frac{\Delta\boldsymbol{v}^2}{V_\mathrm{t}^2}},
\label{Eq:Sc}
\end{equation}
where $\Delta\boldsymbol{v}=\boldsymbol{v}_\mathrm{d}-\boldsymbol{v}_\mathrm{g}$ is the differential velocity between dust and gas. The time-scale for growth typically scales as $(\Omega_\mathrm{K}\,\epsilon)^{-1}\propto r^{\frac{3}{2}}\,\epsilon^{-1}$ \citep{Laibe2008}: dust grains grow more rapidly in the inner disc and when the dust-to-gas ratio is large.

Our simulations show that $\left| \Delta\boldsymbol{v} \right| \ll V_\mathrm{t}$, i.e. Sc is dominated by the effect of gas-dust coupling determined by the value of St. This gives a useful approximation for the variations of the relative velocities:
\begin{equation}
V_\mathrm{rel}\simeq\sqrt{2\tilde{\alpha}}\,\frac{\sqrt{\St}}{1+\St}\,c_\mathrm{s}.
\label{Eq:approxVrel}
\end{equation}
$\Vrel$ remains smaller than
\begin{equation}
\Vrel^\mathrm{max}\equiv\sqrt{\frac{\tilde{\alpha}}{2}}\,c_\mathrm{s},
\label{Eq:Vrelmax}
\end{equation}
which occurs when $\mathrm{St}=1$. In our simulations, $\Vrel^\mathrm{max}\simeq0.15\,c_\mathrm{s}$.

Our dust SPH particles have equal fixed masses and represent dust grains of the same size: with growth, they represent fewer physical grains of larger size (and mass), but still numerous enough to maintain the validity of the numerical scheme \citep{Laibe2008}. The assumption that growth occurs only between grains of the same size has little effect on the global size distribution. Indeed, a given volume contains SPH particles representing grains of different sizes. Our growth model results in a spread of (larger) sizes in that volume. If allowed, growth between grains of different sizes would produce sizes within that same range and only slightly depopulate both ends of the size distribution in that volume.

The spread in the relative velocities between gas and dust is treated in a Lagrangian way by our SPH code. This results in  a spread in $\Vrel$ at a given location in the disc, similarly to the velocity distributions used by \citet{Windmark2012} and \citet{Garaud2013}. The sub-grid description of turbulence is a source of uncertainty of the model. Several expressions for $\Vrel$ have been proposed in the literature, e.g. \citet{Ormel2007} and \citet{Pan2010} --- see \citet{Laibe2014} for a discussion.\footnote{We have run test simulations with several of them, producing qualitatively similar dust behaviors and, importantly, they all lead to a dust pile-up.}

\subsection{Fragmentation model}
\label{sec:Frag}

When two particles collide with high relative kinetic energy, they break into a collection of fragments of various sizes. In a steady state, the size distribution of fragments is commonly represented by a power law.
We implement fragmentation by comparing the relative velocity between grains $\Vrel$ to the fragmentation threshold $\Vfrag$. If $\Vrel>\Vfrag$, grains shatter, which we model by decreasing the size of the representative SPH particles
following
\begin{equation}
\frac{\mathrm{d}s}{\mathrm{d}t}=-\epsilon\,\frac{\rho_\mathrm{g}}{\rho_\mathrm{s}}\,\Vrel.
\label{Eq:dsdt_frag}
\end{equation}
For grains of a given Stokes number, fragmentation occurs preferentially in the inner disc regions, where they have higher relative velocities (see equation~\ref{Eq:approxVrel}, in which $c_\mathrm{s}$ decreases as a function of $r$). Appendix~\ref{app:LimGrowthFrag} gives the initial location of the limit between growth and fragmentation as a function of grain size and distance to the star in our disc models. To ensure consistency with the growth model, as well as to properly conserve physical quantities and avoid a prohibitively large increase in the number of SPH particles, we keep a locally monodisperse size distribution. This means we follow only the largest fragment of the distribution. Equation~\ref{Eq:dsdt_frag} is equivalent to $\mathrm{d}m/\mathrm{d}t=-m/\tau_\mathrm{coll}$ and implies that the largest fragment has lost most of the original grain's mass, i.e. that fragmentation leads to small sizes.
Our model is thus conservative: by producing fragments that are much smaller than their parent grains, it delays their subsequent evolution to larger sizes and makes planetesimal formation more difficult.
For numerical efficiency, we limit the minimum fragmentation size to $s_0=10~\mu$m. This choice has little effect on the overall size distribution. Indeed, very small grains always have small $\Vrel$ --- see equation~\ref{Eq:approxVrel} --- hence are unaffected by fragmentation. They grow very efficiently and quickly forget their initial size \citep{Laibe2008}.

For a range of velocities below $\Vfrag$, colliding grains are expected to bounce instead of sticking or shattering \citep{Zsom2010}, which we do not take into account. This would lower the global growth efficiency, and can be mimicked to first order by reducing $\Vfrag$ (however, this would still produce small fragments, which are not expected in bouncing collisions). We chose to keep only one free parameter and not to include bouncing in this work, leaving an exploration of its effects for a future study. Our treatment is simplified compared to that of \citet{Birnstiel2010} or \citet{Garaud2013}, in particular we do not keep track of the smallest fragments who maintain a population of small grains in the disc that are responsible for the infrared emission, nor do we consider collisions between unequal-sized particles. Yet, the resulting global size distribution we achieve when backreaction is turned off is very similar to that obtained by \citet{Brauer2008} and \citet{Birnstiel2010}, where backreaction was neglected (see Section~\ref{sec:Discussion}).

\section{Results}
\label{sec:Results}

\begin{figure*}
\centering
\resizebox{\hsize}{!}{
\includegraphics{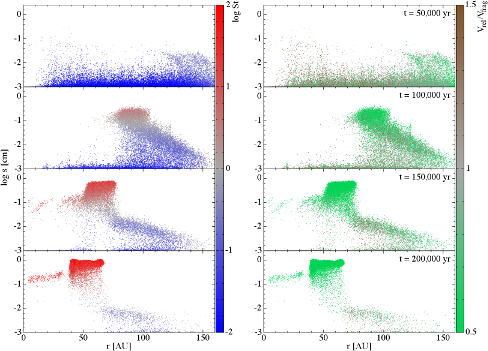}
}
\caption{Radial grain size distribution in the flat disc for $\Vfrag=15$~m\,s$^{-1}$ and $\epsilon_0=1$\%. Grains grow, decouple from the gas and pile up, resulting in a self-induced dust trap. The blue/red colour code represents the Stokes number St (left), and the brown/green colour code represents the ratio $\Vrel/\Vfrag$ (right). Four snapshots at 50,000, 100,000, 150,000 and 200,000~yr are shown, from top to bottom. Video~1 (available online) shows the full evolution.}
\label{Fig:rs-St-Vrel_flat_V15_BR_evol}
\end{figure*}

The evolution of the radial grain size distribution in the flat model with $\Vfrag=15$~m\,s$^{-1}$ and $\epsilon_0=1$\% is displayed in Fig.~\ref{Fig:rs-St-Vrel_flat_V15_BR_evol}. Its full evolution is shown in Video~1 (available online). At early times, growing grains fragment almost immediately in most of the disc, keeping sizes smaller than 100~$\mu$m. Only grains in the very outer disc have relative velocities low enough to avoid fragmentation and grow slowly. After $\sim80,000$~yr, they have reached the size where $\St=1$ ($\sim1$~mm for these disc conditions), drift inwards and continue to grow. At $\sim100,000$~yr, the largest grains begin decoupling from the gas, slow their radial drift and pile up between $\sim70-100$~au. Locally, $\epsilon$ becomes large
, which causes the backreaction to strongly reduce the dust drift velocity, and that is the case for all values of St, including the fastest drift regime when $\St\sim1$ \citep{Nakagawa1986}. This amplifies this pile-up, helping grains escape the radial-drift barrier. To see a similar behavior without backreaction, \citet{Birnstiel2010} had to artificially decrease the drift efficiency by 80\%. The critical point is that backreaction also adds an outwards term to the gas radial velocity, usually negligible compared to the inwards accretion flow, except again when $\epsilon$ is large and $\St\sim~1$ (see Section~\ref{sec:BReffect2} and Appendix~\ref{app:GasDyn} for more details). Near the radial dust concentration, gas is locally dragged outwards, accumulates and creates a pressure maximum. This results in a self-induced dust trap, with no need for any special condition. After 200,000~yr, most of the dust grains are concentrated in a dense ring extending from $\sim40$ to $\sim65$~au, have grown to cm sizes and are well decoupled from the gas, with low relative velocities. For higher values of $\Vfrag$, the dust trap forms closer to the star (see Appendix~\ref{app:Varying} for details).

\begin{figure*}
\centering
\resizebox{\hsize}{!}{
\includegraphics{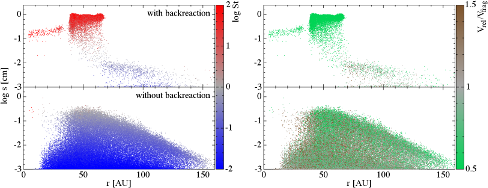}
}
\caption{Top: same as Fig.~\ref{Fig:rs-St-Vrel_flat_V15_BR_evol} after 200,000~yr. Bottom: same simulation with backreaction turned off. In this case, when grains grow and reach St~$\sim1$, they drift rapidly towards the central star while fragmenting and replenishing the small grains reservoir. Video~1 (available online) shows the full evolution.}
\label{Fig:rs-St-Vrel_flat_V15_BR_noBR}
\end{figure*}

We performed the same simulation without the backreaction in order to emphasize the critical role it plays. Figure~\ref{Fig:rs-St-Vrel_flat_V15_BR_noBR} shows that no grain has been able to escape the radial-drift barrier (see Video~1, available online, for the full evolution), similarly to what is found in other studies in which backreaction is neglected, e.g. \citet{Brauer2008,Birnstiel2010,Windmark2012}. Once grains have reached the size for which $\St=1$, they drift inwards to regions where their relative velocities are high and exceed $\Vfrag$. The grains then fragment and replenish the small grains reservoir. The largest fragments keep drifting and are accreted onto the star. No pile-up forms in this case.

\begin{figure}
\centering
\resizebox{\hsize}{!}{
\includegraphics{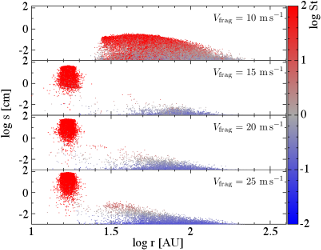}
}
\caption{Radial grain size distribution in the steep disc after 400,000~yr, for $\epsilon_0=1\%$ and different values of the fragmentation velocity. From top to bottom: $\Vfrag=10$, 15, 20 and 25~m\,s$^{-1}$. For $\Vfrag=10$~m\,s$^{-1}$, a self-induced dust trap forms at $r\sim30$~au, at the fragmentation front of the particles where $\Vrel(\St=1)=\Vfrag$. For $\Vfrag\ge15$~m\,s$^{-1}$, the trap forms at \mbox{$r\sim15$--20~au}, where growth induces a strong pile-up. A log scale has been used to emphasize these different locations.}
\label{Fig:rs-St_steep_e1}
\end{figure}

\begin{figure}
\centering
\resizebox{\hsize}{!}{
\includegraphics{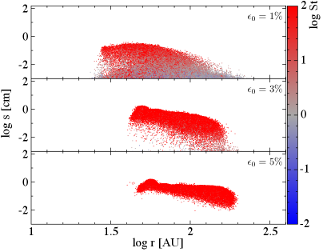}
}
\caption{Radial grain size distribution in the steep disc after 400,000~yr, for $\Vfrag=10$~m\,s$^{-1}$ and different values of the initial dust-to-gas ratio. From top to bottom: $\epsilon_0=1$, 3 and 5\%. Larger grain concentrations give rise to more effective dust trapping.}
\label{Fig:rs-St_steep_V10}
\end{figure}

Self-induced dust traps also develop in the steep disc for a range of conditions. For $\epsilon_0=1$\% and $\Vfrag=10$~m\,s$^{-1}$, a trap forms at $\sim30$~au (top panel of Fig.~\ref{Fig:rs-St_steep_e1}). Behind it, the region of large, decoupled grains extends out to $\sim100$~au because of the different disc structure. For higher values of $\Vfrag$, fragmentation is less of a hindrance and grains can grow efficiently at smaller radial distances where the growth time-scale, varying as $r^\frac{3}{2}$ \citep{Laibe2008}, is shortest. This behavior is similar to the case where fragmentation is not included (equivalent to $\Vfrag=+\infty$) and grains pile up very efficiently in the inner disc, as predicted theoretically \citep{Laibe2014} and seen in simulations \citep{Gonzalez2015a}, at the same location for all values of $\Vfrag\ge15$~m\,s$^{-1}$ (bottom three panels of Fig.~\ref{Fig:rs-St_steep_e1}). The gas then accumulates and the self-induced dust trap forms. Note that this radial concentration is not caused by the inner boundary: grains initially in the inner disc drift inwards and leave the disc before dust piles up at 15--20~au (see Video~2, available online, which shows the evolution of radial grain size distribution in the steep disc with $\Vfrag=15$~m\,s$^{-1}$ and $\epsilon_0=1$\%, with and without backreaction). For higher values of $\epsilon_0$, grains grow faster (equation~\ref{Eq:dsdt}), decouple sooner during their inward drift, and therefore pile up at larger radii --- see Fig.~\ref{Fig:rs-St_steep_V10}. Results for the full range of values of $\epsilon_0$ and $\Vfrag$ are detailed in Appendix~\ref{app:Varying}.

\section{Discussion}
\label{sec:Discussion}

The formation mechanism of the self-induced dust trap is summarized in the sketch presented in Fig.~\ref{Fig:sketch}, and detailed in the following subsections.

\subsection{Without backreaction: radial-drift and fragmentation barriers}
\label{sec:noBR}

\begin{figure}
\centering
\resizebox{\hsize}{!}{
\includegraphics{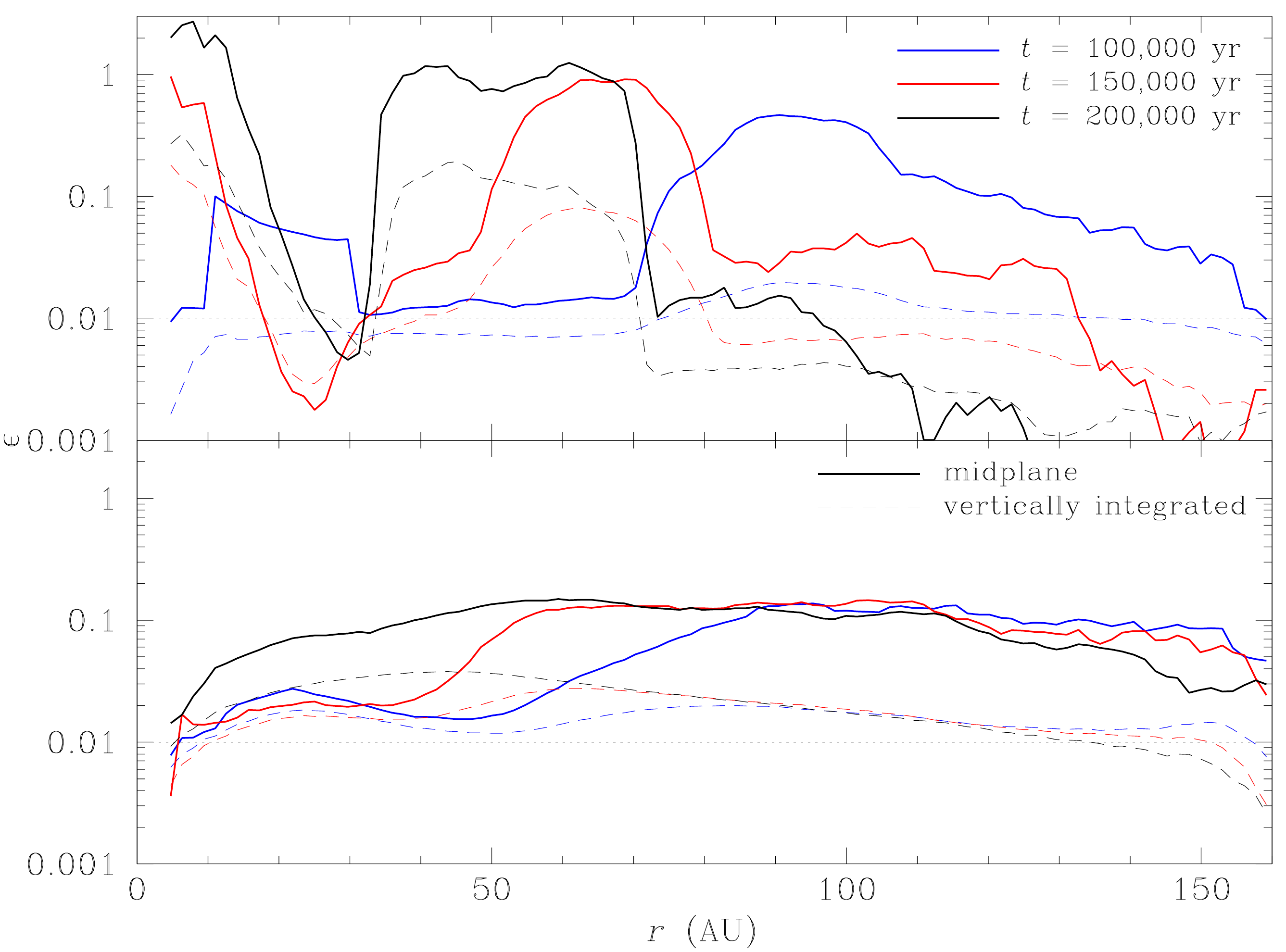}
}
\caption{Radial profiles of the dust-to-gas ratio $\epsilon$ in the flat disc with an initial value $\epsilon_0=1$\% (marked by the horizontal dotted line), for $\Vfrag=15$~m\,s$^{-1}$. Its midplane and vertically integrated values are shown at 100,000, 150,000 and 200,000~yr for the simulations with (top) and without (bottom) backreaction. The dust-to-gas ratio is enhanced in the midplane because of vertical dust settling. All radial profiles shown in this figure as well as in Figs.~\ref{Fig:barriers_flat_V15_noBR}, \ref{Fig:eps_Sr_x_br_flat_V15}, \ref{Fig:Pmid_flat_V15}, \ref{Fig:gasflux_flat_V15}, \ref{Fig:Pmid_steep_e1_3_5_V10}, \ref{Fig:Pmid_flat_V15_20_25} and \ref{Fig:Pmid_steep_e1_5_V10_15_20_25} are azimuthally averaged.}
\label{Fig:epsilon_flat_V15}
\end{figure}

\begin{figure}
\centering
\resizebox{\hsize}{!}{
\includegraphics{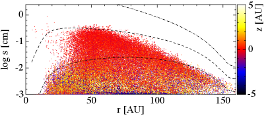}
}
\caption{Radial grain size distribution in the flat disc with $\Vfrag=15$~m\,s$^{-1}$, $\epsilon_0=1$\% without backreaction after 200\,000~yr, colour-coded by altitude $z$. Radial profiles of the sizes $s_\mathrm{drift}$ (dashed line) and $s_\mathrm{frag}^\pm$ (dash-dot lines) are overplotted.}
\label{Fig:barriers_flat_V15_noBR}
\end{figure}

The first simulations of the dust evolution including fragmentation were performed by \citet{Brauer2008} and \citet{Birnstiel2010} in 1D along the radial direction on a fixed, vertically and azimuthally averaged gas disc structure, and neglecting backreaction. Their results are qualitatively reproduced in our 3D simulations with backreaction turned off, in which $\Vfrag=15$~m\,s$^{-1}$ and $\epsilon_0=1\%$ (Fig.~\ref{Fig:rs-St-Vrel_flat_V15_BR_noBR} and bottom panels of Video~1 for the flat disc and of Video~2 for the steep disc). In the inner disc, growing grains re-fragment rapidly and repopulate the reservoir of small grains, which drift very slowly. Grains growing in the outer disc reach sizes close to $s_\mathrm{drift}$ and are dragged towards the interior where they fragment to small sizes or are lost from the disc (see the bottom-right panels of Videos~1 and 2). Grains are not able to grow above $s_\mathrm{drift}$, and hence they do not break the radial-drift barrier. This was the main result of previous studies neglecting backreaction \citep{Brauer2008,Birnstiel2010}. The dust-to-gas ratio shows an inward moving front, illustrating the dust drift (shown for the flat disc in the bottom panel of Fig.~\ref{Fig:epsilon_flat_V15}). In the midplane, $\epsilon$ is larger than its vertical average, due to dust settling, and reaches about 10\%. This ten-fold increase is consistent with the thickness of the dust layer predicted by \citet{Dubrulle1995} for $\alpha=10^{-2}$. Figure~\ref{Fig:barriers_flat_V15_noBR} shows the final radial grain size distribution in the flat disc, colour-coded by altitude in the disc. It shows that the bulk of the grain population is close to the midplane and is drift-limited (the grain sizes stay below $s_\mathrm{drift}$). In addition, Fig.~\ref{Fig:barriers_flat_V15_noBR} shows that the grains closer to the disc surface have sizes below the fragmentation size $s_\mathrm{frag}^-$ (see Appendix~\ref{app:LimGrowthFrag} for its derivation), i.e. that they are fragmentation-limited, something 1D simulations cannot see. Such simulations would also only notice a moderate increase in the vertically averaged dust-to-gas ratio (Fig.~\ref{Fig:epsilon_flat_V15}).

\subsection{Backreaction effect \#1: Radial concentration of grains}
\label{sec:BReffect1}

When backreaction is considered, the dust drift velocity is reduced by a factor $(1+\epsilon)^2$, for all values of $\St\lesssim1$ \citep{Nakagawa1986}. Negligible at first, this effect becomes important when dust settles and $\epsilon$ increases in the midplane. This results in a radically different dust evolution compared to simulations without backreaction (Fig.~\ref{Fig:rs-St-Vrel_flat_V15_BR_evol} and top panels of Video~1 for the flat disc and of Video~2 for the steep disc, for $\Vfrag=15$~m\,s$^{-1}$). Initially, fragmentation dominates the dust evolution in the inner disc, similarly to the case without backreaction. However, grains exterior to $r_\mathrm{frag}$ grow unhindered by fragmentation and backreaction acts to slow down their drift. This helps them to grow above $s_\mathrm{frag}^+$ (the top-right panels of Videos~1 and 2 show that they have $\Vrel<\Vfrag$) before entering the inner fragmentation region (see Appendix~\ref{app:LimGrowthFrag}). They progressively decouple from the gas and slow down. In the flat disc, after $\sim100,000$~yr, dust has piled up and formed a radial concentration of  marginally coupled dust grains at $\sim90$~au, as can be seen in the radial profile of $\epsilon$ in the midplane in Fig.~\ref{Fig:epsilon_flat_V15}. (Note that the large values of $\epsilon$ interior to 20--30~au are due to a decrease of the gas density rather than an increase of the dust density.) The dust concentration drifts slowly and keeps decelerating, helped by the increasing importance of backreaction when $\epsilon$ approaches unity. The drift stalls, with the inner edge of the dust concentration around 40~au.

\subsection{Backreaction effect \#2: Dust concentrations modify the gas structure}
\label{sec:BReffect2}

The first stage of the self-induced dust trapping mechanism consists in the formation of the strongly peaked radial concentration of dust grains with $\St\sim1$, as just described. We now explain the second stage, i.e. how dust backreaction modifies the gas structure around the dust concentration. In Appendix~\ref{app:GasDyn}, we give the analytical expression of the gas radial velocity as the sum of the viscous inwards flow and the outwards motion due to dust drag. In most situations, the viscous flow dominates, but in dust concentrations $\epsilon$ becomes large and the outwards drag locally becomes the dominant term. To demonstrate the effect on the gas structure, we numerically integrate the evolution equation of the gas surface density in the presence of a sharp dust concentration. We observe an outwards mass transfer and a gas accumulation at the outer edge of the dust concentration (see Fig.~\ref{Fig:maxgas}), in agreement with the analytical prediction. The gas evolution equation (equation~\ref{eq:approx}) stresses the role played by the following three key ingredients: i) backreaction, whose effect on the gas phase is represented by the $f_\mathrm{drag}$ term (defined in Appendix~\ref{app:GasDyn}), ii) growth and fragmentation, assisted by backreaction, which produce local dust concentrations with $\epsilon \simeq \mathrm{St} \simeq 1$, hence large values of $f_\mathrm{drag}$, and iii) global dust dynamics, giving rise to relevant large-scale gradients of $f_\mathrm{drag}$. This mechanism can not be simulated using local approximations.

\begin{figure}
\centering
\resizebox{\hsize}{!}{
\includegraphics{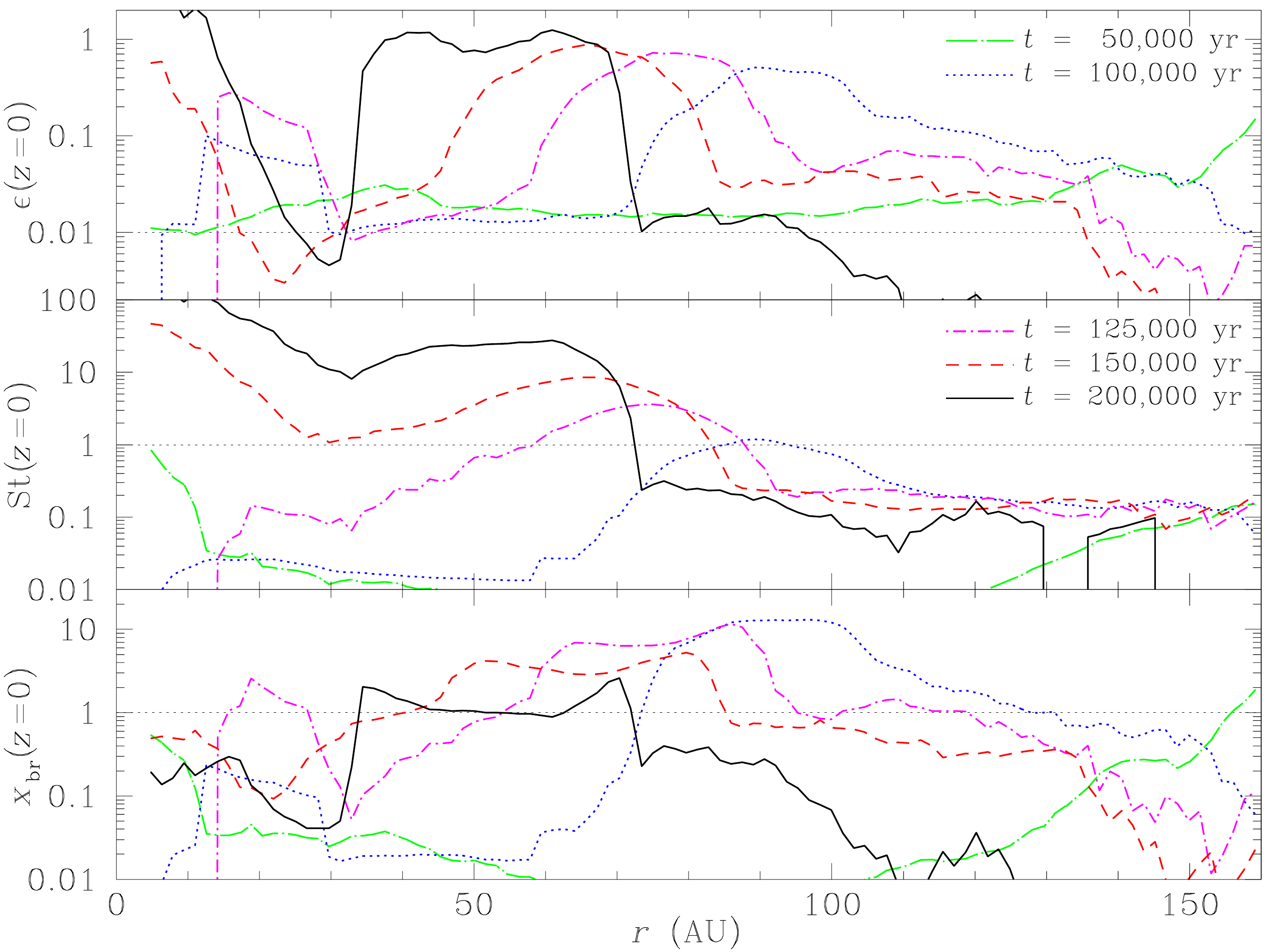}
}
\caption{Evolution of $\epsilon$, $\St$ and $x_\mathrm{br}$ in the midplane of the flat disc with $\Vfrag=15$~m\,s$^{-1}$ and $\epsilon_0=1$\%.}
\label{Fig:eps_Sr_x_br_flat_V15}
\end{figure}

\begin{figure}
\centering
\resizebox{\hsize}{!}{
\includegraphics{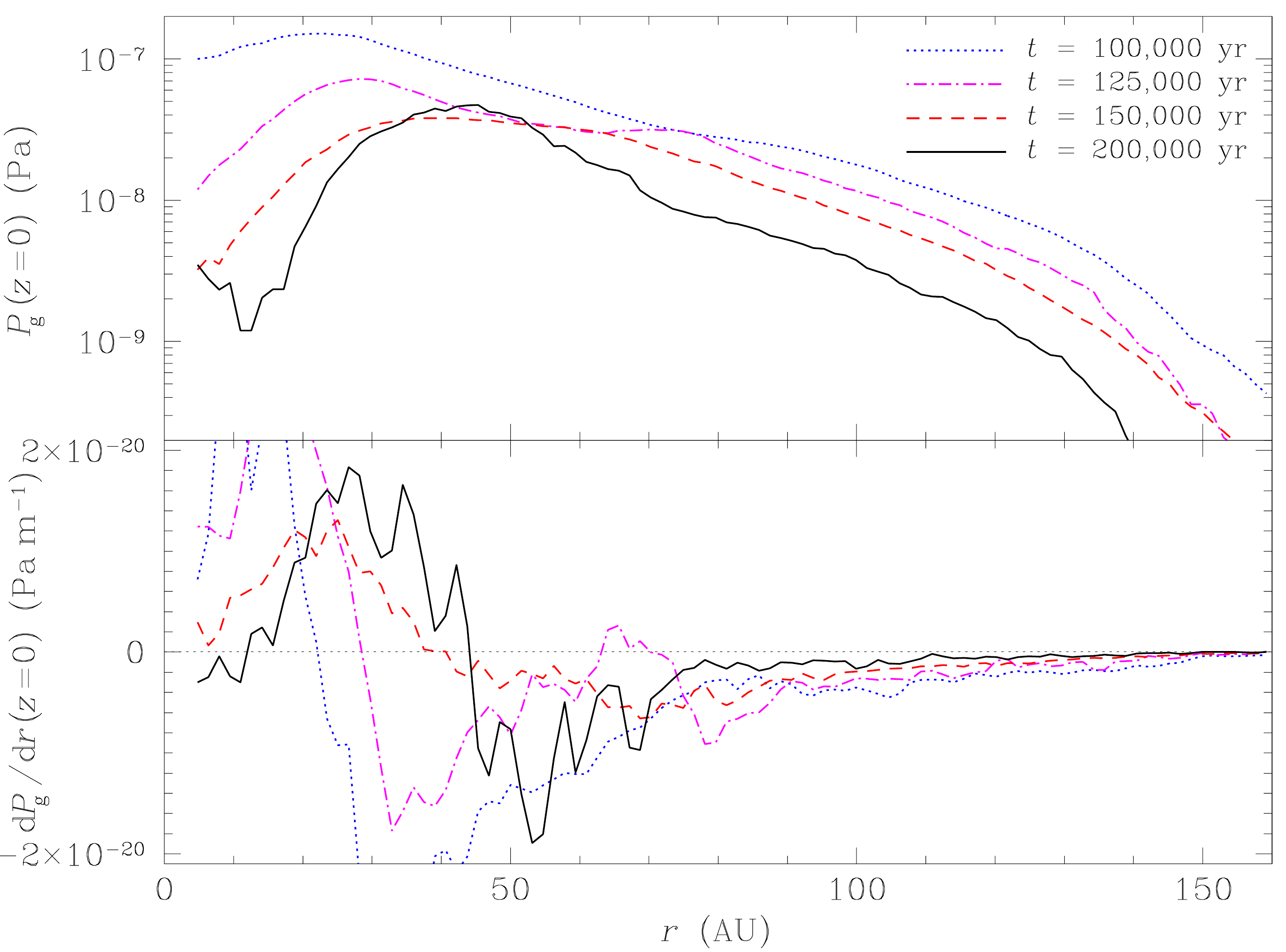}
}
\caption{Evolution of the gas pressure (top) and pressure gradient (bottom) in the midplane of the flat disc with $\Vfrag=15$~m\,s$^{-1}$ and $\epsilon_0=1$\%.}
\label{Fig:Pmid_flat_V15}
\end{figure}

\begin{figure}
\centering
\resizebox{\hsize}{!}{
\includegraphics{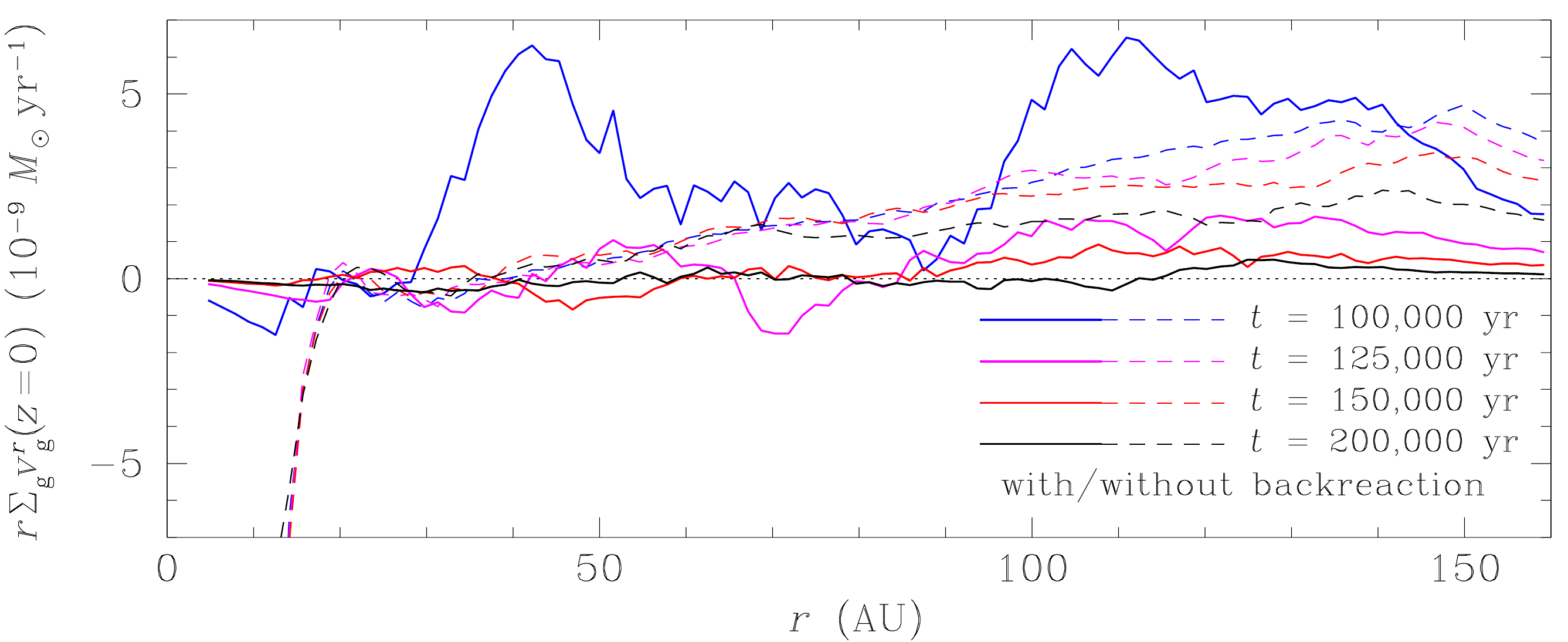}
}
\caption{Evolution of the radial density flux of the gas in the midplane of the flat disc with $\Vfrag=15$~m\,s$^{-1}$ and $\epsilon_0=1$\% for the simulations with and without backreaction.}
\label{Fig:gasflux_flat_V15}
\end{figure}

Quantitatively, the relative importance between backreaction and viscosity is measured by the parameter $x_\mathrm{br}$, a function of $\epsilon$ and $\St$ defined in equation~\ref{eq:def_xbr}. As an example, we discuss the role played by backreaction in the case of the flat disc with $\Vfrag=15$~m\,s$^{-1}$ in Fig.~\ref{Fig:eps_Sr_x_br_flat_V15}. Before 100,000~yr, backreaction is negligible ($x_\mathrm{br}\ll1$ since $\epsilon$ and St are small). However, around 100,000~yr, the motion of the gas is driven by backreaction since both $\epsilon$ and St approach unity and $x_\mathrm{br}\gg1$. The resulting outwards flow starts to modify the gas pressure profile, causing a slight bump at $\sim90$~au shown in Fig.~\ref{Fig:Pmid_flat_V15}. At later stages, the disc region affected by backreaction has drifted inwards, together with the radial concentration of slowly decoupling grains (Fig.~\ref{Fig:eps_Sr_x_br_flat_V15}, $x_\mathrm{br}$ decreases but remains well above unity). Figure~\ref{Fig:Pmid_flat_V15} shows the formation of a pressure maximum, which has become stationary by 200,000~yr, when the grains in the radial concentration have fully decoupled (see Fig.~\ref{Fig:rs-St-Vrel_flat_V15_BR_evol}), and centered around 40~au. At this stage, the viscous and drag terms balance each other (Fig.~\ref{Fig:eps_Sr_x_br_flat_V15} shows that $x_\mathrm{br}\sim1$ at the location of the pressure maximum).

The motion of the gas phase is illustrated by the profile of its radial density flux, plotted in Fig.~\ref{Fig:gasflux_flat_V15}. At 100,000 ~yr, it shows outwards mass transport exterior to $\sim30$~au and vanishing at $\sim90$~au where gas thus accumulates in a nascent pressure bump (Fig.~\ref{Fig:Pmid_flat_V15}). Outwards motion is seen again further out as the disc viscously spreads. At 125,000 and 150,000~yr, the flux profiles exhibit adjacent regions of outwards and inwards mass transport, i.e. converging motion towards 65 and 40~au, respectively, strengthening the pressure maximum seen in Fig.~\ref{Fig:Pmid_flat_V15}. At the end of the simulation, no gas motion is seen when the pressure maximum has fully formed and is stationary. As expected, no such behaviour is observed in the simulation without backreaction, which shows only the usual accretion in the inner disc and viscous spreading in the outer disc.

A dust trap has thus formed at 40~au, coincident with the inner edge of the dust accumulation seen in Figs.~\ref{Fig:rs-St-Vrel_flat_V15_BR_evol} and \ref{Fig:epsilon_flat_V15} at the end of the simulation. Note that even though the gas accumulates at the outer edge of the initial (and narrow) dust concentration (Appendix~\ref{app:GasDyn}), inward drifting grains are subsequently stopped by the gas pressure maximum and thus pile up exterior to it, forming a wider dust accumulation.

\subsection{Self-induced dust traps}
\label{sec:SIDT}

The combined evolution of the gas and dust phases detailed above leads to the formation of a gas pressure maximum, which acts as a dust trap. This trap does not result from a pre-existing discontinuity in the disc, such as a dead zone edge or a snow line or from the opening a planet gap, but forms spontaneously --- a self-induced dust trap. It then acts as any other dust trap and stops the inward motion of grains drifting from the outer disc, which accumulates them further. Trapped grains have sufficiently low collision velocities to stick and continue growing. At the end of our simulations, the mass of trapped dust ranges from a few to a few tens of Earth masses (with higher values for larger $\Vfrag$ and larger $\epsilon_0$). Enough solid mass has accumulated to allow the formation of planetesimals. Modelling their actual formation would require self-gravity, which is not included in our simulations, and is out of the scope of this paper.

Backreaction plays a notable role at every step of the evolution. First, it enhances the radial concentration of dust grains by slowing down their drift, which increases the dust-to-gas ratio $\epsilon$ and accelerates grain growth, which in turn slows the particles even more. Second, it reshapes the gas structure by transferring gas from the inner to the outer regions. A pressure maximum forms, preventing further drift of the particles towards the star.

In summary, without backreaction the dust population stays drift- or fragmentation-limited, as commonly seen by other authors. It is only when the self-consistent evolution of both gas and dust is fully taken into account, with backreaction included, that a different behaviour emerges: drifting, growing and fragmenting grains are able to break the radial-drift and fragmentation barriers and form a stationary radial concentration.

This self-induced trapping mechanism is robust. Whatever the disc conditions, fragmentation threshold, or initial dust-to-gas ratio, dust traps form, but at different locations and different evolutionary times. The larger the dust mass, the stronger the effect. In particular, higher dust-to-gas ratios cause grains to pile up at larger radii, as well as grow to larger sizes. If they end up forming planetesimals, this could explain the presence of planets orbiting several tens of au from their stars: a value of $\epsilon_0$ slightly larger than average is enough for pebbles to accumulate in the outer disc.

Note that, since grains in our self-induced dust traps have reached mm- to cm-sizes, they would appear as bright rings in dust continuum observations in the mm range (see \citealt{Gonzalez2015b} for ALMA simulated images of such a dust concentration). At later times, if planetesimals form at the trap location, a deficit of emission would be seen as a concentric gap.

\subsection{Comparison with the streaming instability}
\label{sec:SI}

We aim to compare the formation of the self-induced dust traps identified above (defined by the equations given in Appendix~\ref{app:GasDyn}) to the streaming instability, discovered and defined by the set of equations given in \citet{Youdin2005}. Both mechanisms are particular cases of two-stream instabilities in discs, which also encompass the generation of non-axisymmetric structures by dust backreaction \citep{Lyra2013}. Both of these instabilities require a dust-and-gas mixture in rotation, and an additional background pressure gradient in the gas.

The streaming instability develops in the $(r,z)$ plane of a disc whose background pressure gradient is enforced to remain constant. The transient linear growth stage has been studied analytically by \citet{Youdin2005,Youdin2007,Jacquet2011} and shows efficient local enhancement of the dust density. The non-linear regime has been investigated numerically \citep{Johansen2007,Bai2010,Yang2014} and leads to strong dust clumping, assisted by the slowdown of radial drift in overdense regions. The streaming instability is a proposed mechanism to overcome the barriers of planet formation and lead to the formation of planetesimals once self-gravity becomes important \citep{Johansen2007}. The self-induced dust traps we describe in this work also lead to strong concentrations of dust grains, which no longer drift and can grow unhindered by fragmentation. They provide therefore a solution to the barriers of planet formation as well. The two mechanisms share similar physical ingredients, but we stress their differences below since they have important consequences for planet formation.

The streaming instability can grow even in the limit of a rigorously incompressible gas \citep{Youdin2005}, i.e. gas compression is not required for the local dust-to-gas ratio to be amplified. The perturbation must develop in the radial but also in the vertical direction to grow (i.e., the flow is not unstable if the wavenumber $k_z=0$, e.g. see equations~29 and 32 of \citealt{Jacquet2011} in the terminal velocity approximation). Self-induced dust traps require large scale background gradients for the dust-to-gas ratio and the Stokes number, assisted by an efficient compression of the gas in the midplane to develop, i.e. the formation of the pressure maximum is triggered by the term on the right-hand side of equation~\ref{eq:approx}, which comes from the compressibility term $\Sigma_\mathrm{g} \boldsymbol{\nabla \cdot v} \ne 0$. The instability develops along the sole radial direction. Global simulations of evolved dust distributions with backreaction are therefore needed to observe the formation of self-induced dust traps.

The time-scale for the streaming instability to develop under the terminal velocity approximation, when $\St\ll1$ is $\mathcal{O}\left((r/H)^{2}t_\mathrm{K}\right)$, see \citet{Youdin2005}. In simulations, it is found to be $\sim10$--$40\,t_\mathrm{K}$ \citep{Johansen2007b,Bai2010,Yang2014}. This is somewhat longer than the typical time required for the self-induced dust trap to form, $\tau\sim5\,t_\mathrm{K}$ (see equation~\ref{eq:tau}). This suggests that where \textit{large-scale} radial dust concentrations build up, self-induced dust traps should form, and that \textit{small-scale} streaming instability may develop over longer time-scales. Concurrent appearance of both can not be ruled out. The streaming instability and self-induced dust traps both lead to the formation of dust clumps, but along different paths. The streaming instability starts to concentrate solids as early as in its transient linear growth stage. Self-induced dust traps require first a local enhancement of the dust-to-gas ratio by settling or radial pile-up. Then, a pressure maximum forms, without affecting the local dust density. Dust is then collected by this trap over longer time-scales, as particles drifting from the outer disc regions accumulate in the pressure maximum.

Finally, the streaming instability requires low viscosities for the perturbation to grow without being damped by viscous dissipation. Equation 33 of \citet{Youdin2005} gives a maximum value of
\begin{equation}
\alpha \la \frac{g}{\Omega_\mathrm{K}} \left(\frac{H}{r} \frac{2\pi}{K}\right)^2,
\end{equation}
where $g$ denotes the linear growth rate of the instability and $K$ the normalised wavenumber. In the favorable case where $\St\sim1$ and $\epsilon\sim0.1$--1, \citet{Youdin2007} find $g/\Omega_\mathrm{K}\sim0.1$ and $K\sim1$, which gives, with $H/r\sim0.05$, a maximum value $\alpha \sim10^{-2}$, comparable to the values in our discs. These conditions are however found only transitorily in our simulations and the streaming instability would not be operative in our large viscosity discs when $\St$ and $\epsilon$ have different values. For example, for $\St\sim0.1$ and $\epsilon\sim0.1$--1, the streaming instability develops only when $\alpha\la2$--$8\times10^{-5}$. Self-induced dust traps, on the other hand, have no problem forming with large viscosities: backreaction dominates the gas motion for a wide range of $\St$ and $\epsilon$ values when $\alpha=10^{-2}$ --- see Fig.~\ref{Fig:x_br} --- and for a reduced range for $\alpha$ as high as $10^{-1}$ (equation~\ref{eq:def_xbr}). We note that the simulations presented here lack the necessary resolution to pick up the small-scale fastest-growing mode of the streaming instability. This is of no consequence in this work, since the viscosity is large enough for the streaming instability to operate only marginally.

Thus, although the streaming instability and self-induced dust traps involve some common physical processes, they are two distinct mechanisms with different outcomes (see also Sect.~\ref{sec:pred}). We recommend to refer to these two mechanisms by two distinct names.

\subsection{Where do planetesimals form?}
 \label{sec:pred}
 
The expected location and number of planetesimals are tightly related to the underlying formation mechanism. Self-induced dust traps favour an effective concentrations of solids, but this concerns only a small number of rings located at a few tens of AU from the central star, at most. If planetesimals form inside the pressure bump by gravity in a few orbits, a significant population of observable grains should remain preserved outside of the trap. If planetesimals form in low-viscosity discs by the streaming instability, there is no physical reason to expect the formation to be local: the conditions for the instability to develop should be fulfilled in extended regions of the disc, if not everywhere. Moreover, numerical simulations show that once activated, it is a powerful mechanism to convert grains into planetesimals \citep{Johansen2007,Bai2010}. Hence, planetesimal formation by the streaming instability should be global and total, preserving only a tiny population of grains in regions where it develops. Thus, self-induced dust traps and the streaming instability provide two potential paths towards planetesimal formation. Statistics over future spatially resolved observations of young discs might help to determine the relevance of each of them.

\section{Conclusion}
\label{sec:Conclusion}

The core-accretion paradigm of planet formation requires the concentration and growth of primordial grains, a mechanism assumed to be hampered by the radial-drift and fragmentation barriers.  In this work we show that both of these barriers to planet formation can be overcome by a self-induced dust trap, which requires no special conditions or new physics.  The key ingredients are the self-consistent inclusion of the backreaction of the dust onto the gas due to aerodynamic drag, which becomes important when the dust-to-gas ratio locally approaches unity, along with growth and fragmentation of dust grains. In global disc simulations of dust dynamics, growing grains that have settled to the disc midplane progressively decouple from the gas and slow their radial migration. The dust piles up and these radial dust concentrations modify the gas structure on large scales due to the backreaction, producing a gas pressure maximum that acts as a dust trap. The low collision velocities in the dust trap further enhances grain growth, rapidly leading to pebble-sized grains. The subsequent growth of these pebbles to planetesimals can progress via either collisions or gravitational collapse as demonstrated by \citet{Johansen2014} and \citet{Levison2015}.

We demonstrate that this process is extremely robust and that self-induced dust traps form in different disc structures, with different fragmentation thresholds, and for a variety of initial dust-to-gas ratios.  Changing these parameters result in self-induced dust traps at different locations in the disc, and at different evolutionary times.  While seemingly counter-intuitive, fragmentation is a vital ingredient for planet formation as it helps to form dust traps at large distances from the star. Indeed, fragmentation only allows grains to grow exterior to a certain radial distance and when grains decouple from the gas and start piling up, they do so near that radius. Stronger fragmentation, with a lower fragmentation threshold, implies that this radius lies farther away from the star. This would suggest that most discs thus retain and concentrate their grains at specific locations in time-scales compatible with recent observations of structures in young stellar objects \citep{HLTau2015,Andrews2016}. 

The critical role of the backreaction is demonstrated by comparing with simulations in which it is not included. As seen in previous work, without backreaction a gas pressure maximum does not form and hence grains continue drifting inwards toward the star.  Grains with relative large velocities cannot overcome the fragmentation barrier, fragmenting upon collision and replenishing the small grain population, while the largest grains cannot overcome the radial-drift barrier and are rapidly accreted into the central star.

The self-inducted dust trap presented here is not the same mechanism as the streaming instability. The differences between both have consequences for planetesimal formation: (i) self-induced dust traps develop on somewhat shorter time-scales than the streaming instability, ii) they can develop even in highly viscous discs ($\alpha \simeq 10^{-2} - 10^{-1}$), and iii) they promote planetesimal  formation in a small number of rings while the streaming instability favours large regions of the disc. It is likely, however, that both play a part in planet formation.

In this work, we have presented a powerful new mechanism to overcome the main bottleneck of planet formation --- the growth from micrometre-sized grains to centimetre-sized and decimetre-sized pebbles. This vital phase of grain growth is mostly hindered by the radial-drift barrier and the fragmentation barrier. Once grains have reached pebble sizes, there are several possible pathways for their subsequent growth to planetesimals. Thus self-induced dust traps bring a critical missing piece of the planet formation puzzle and provide a favorable environment for the growth towards planetesimals.

\section*{Acknowledgements}

We thank the anonymous referee for thorough comments that helped us to clarify and improve the manuscript.
This research was partially supported by the Programme National de Physique Stellaire and the Programme National de Plan\'etologie of CNRS/INSU, France. JFG thanks the LABEX Lyon Institute of Origins (ANR-10-LABX-0066) of the Universit\'e de Lyon for its financial support within the programme `Investissements d'Avenir' (ANR-11-IDEX-0007) of the French government operated by the ANR. GL is grateful for funding from the European Research Council for the FP7 ERC advanced grant project ECOGAL. STM acknowledges partial support from PALSE (Programme Avenir Lyon Saint-Etienne). Simulations were run at the Common Computing Facility (CCF) of LABEX LIO. Figures~\ref{Fig:rs-St-Vrel_flat_V15_BR_evol}, \ref{Fig:rs-St-Vrel_flat_V15_BR_noBR}, \ref{Fig:rs-St_steep_e1}, \ref{Fig:rs-St_steep_V10} and \ref{Fig:barriers_flat_V15_noBR} were made with SPLASH \citep{Price2007}.



\bibliographystyle{mnras}
\bibliography{SelfTraps}


\appendix
\section{Growth vs.\ fragmentation}
\label{app:LimGrowthFrag}

At a given location in the disc, the grain size separating the ranges of growing and fragmenting grains is estimated by writing $\Vrel=\Vfrag$ in combination with equation~\ref{Eq:approxVrel}. This gives a second order equation for St:
\begin{equation}
\St^2+2\left(1-\frac{\tilde{\alpha}\,c_\mathrm{s}^2}{\Vfrag^2}\right)\St+1=0.
\label{Eq:St_2nd_order}
\end{equation}
When $\Vfrag>\Vrel^\mathrm{max}$, equation~\ref{Eq:St_2nd_order} has no solution: $\Vrel<\Vfrag$ for all sizes and grains always grow. From equation~\ref{Eq:Vrelmax}, this occurs exterior to a radius
\begin{equation}
r_\mathrm{frag}=r_0\left(\frac{\tilde{\alpha}\,c_\mathrm{s0}^2}{2\Vfrag^2}\right)^\frac{1}{q},
\label{Eq:r_frag}
\end{equation}
where the disc is colder and the turbulence less effective, resulting in low relative velocities between dust grains.
Otherwise, the solutions to equation~\ref{Eq:St_2nd_order} can be translated into limiting sizes for fragmentation
\begin{equation}
s_\mathrm{frag}^\pm =s_\mathrm{drift}\left[\frac{\tilde{\alpha}\,c_\mathrm{s}^2}{\Vfrag^2}-1\pm\frac{\sqrt{\tilde{\alpha}}\,c_\mathrm{s}}{\Vfrag}\sqrt{\frac{\tilde{\alpha}\,c_\mathrm{s}^2}{\Vfrag^2}-2}\right].
\label{Eq:s_frag}
\end{equation}
Grains grow if $s<s_\mathrm{frag}^-$ or $s>s_\mathrm{frag}^+$ and fragment if $s_\mathrm{frag}^-<s<s_\mathrm{frag}^+$.

\begin{figure}
\centering
\resizebox{\hsize}{!}{
\includegraphics{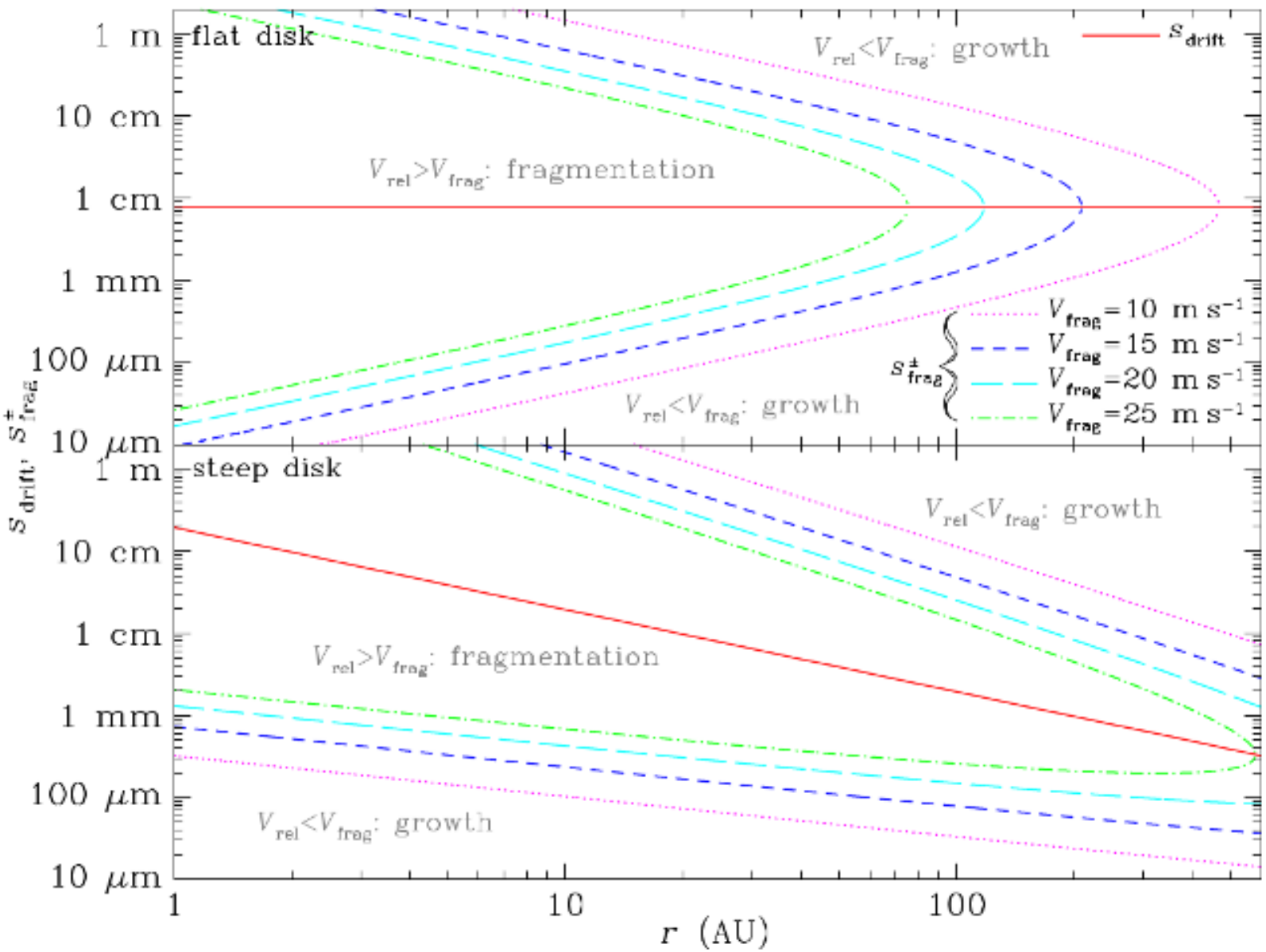}
}
\caption{Initial radial profiles of the sizes $s_\mathrm{drift}$ and $s_\mathrm{frag}^\pm$ in the power-law setup of the flat (top) and steep (bottom) discs. Between the branches $s_\mathrm{frag}^-$ and $s_\mathrm{frag}^+$, $\Vrel>\Vfrag$ and grains fragment, while they grow below $s_\mathrm{frag}^-$ and above $s_\mathrm{frag}^+$. Exterior to a radius $r_\mathrm{frag}$ (which depends on $\Vfrag$), grains always grow.}
\label{Fig:sfrag}
\end{figure}

The radial profiles of both limiting sizes $s_\mathrm{frag}^-$ and $s_\mathrm{frag}^+$, and of the size of fastest drift $s_\mathrm{drift}$, are plotted in Fig.~\ref{Fig:sfrag} for the \textit{initial power-law setup} of the flat and steep discs. At early times, grains are small and grow. In the inner regions of both discs, grains hit the fragmentation barrier at sizes $s_\mathrm{frag}^-$ which are well below $s_\mathrm{drift}$, thus keeping $\St\ll1$ and stay small and well coupled to the gas. They drift very little and remain in the disc. Exterior to $r_\mathrm{frag}$, grains grow unhampered by fragmentation and start drifting rapidly as their size approaches $s_\mathrm{drift}$. If they drift faster than they grow, they will reach the fragmentation barrier, shatter to small sizes and stop drifting. Otherwise, they stay above $s_\mathrm{frag}^+$, keep growing and progressively reach large $\St$ values, decouple from the gas and drift more and more slowly. Note that the profiles of $s_\mathrm{drift}$ and $s_\mathrm{frag}^\pm$ evolve with time, following the evolution of the gas 
density (see equation~\ref{Eq:s_drift}). In particular, 
$\rho_\mathrm{g}$ decreases in the outer disc due to viscous spreading, which shifts the profiles to lower sizes and allows more grains to have $s>s_\mathrm{frag}^+$.

\section{Initial dust-to-gas ratio and fragmentation threshold}
\label{app:Varying}

\begin{figure}
\centering
\resizebox{\hsize}{!}{
\includegraphics{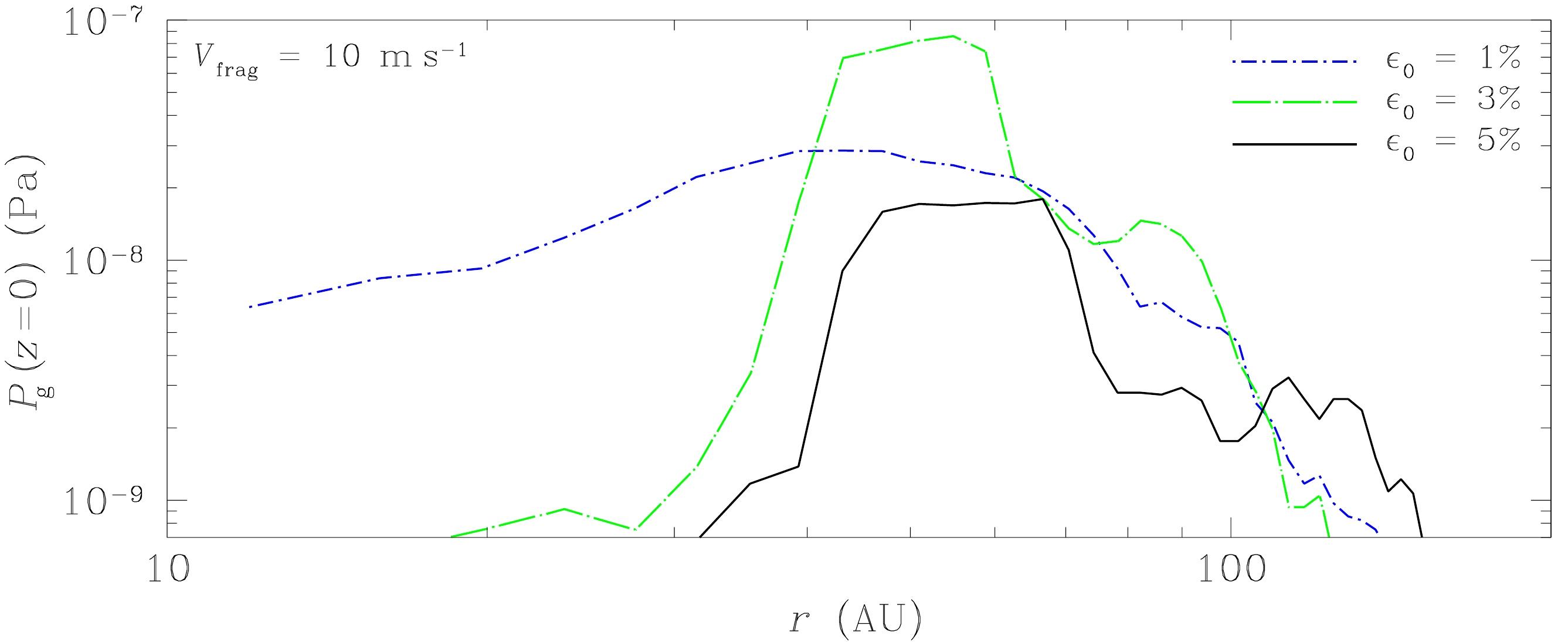}
}
\caption{Radial profiles of the gas pressure in the midplane of the steep disc with $\Vfrag=10$~m\,s$^{-1}$ after 400,000~yr for $\epsilon_0=1$, 3 and 5\%.}
\label{Fig:Pmid_steep_e1_3_5_V10}
\end{figure}

The dust growth rate $\mathrm{d}s/\mathrm{d}t$ is proportional to the dust-to-gas ratio $\epsilon$ (equation~\ref{Eq:dsdt}) and the drift velocity $\mathrm{d}r/\mathrm{d}t$ is a decreasing function of $\epsilon$ \citep{Nakagawa1986}. Therefore, $\mathrm{d}s/\mathrm{d}r=(\mathrm{d}s/\mathrm{d}t)/(\mathrm{d}r/\mathrm{d}t)$ is an increasing function of $\epsilon$. In discs with larger $\epsilon_0$, grains grow faster, and when they decouple from the gas and stop drifting, they are farther away from the star (in the $r-s$ plane, the slope of a grain's trajectory is steeper, similarly to what is seen in the top panel of Fig.~\ref{Fig:sketch} in the $r-\St$ plane, comparing the cases without and with backreaction). The self-induced dust traps form sooner and at larger radial distances, as can be seen in Fig.~\ref{Fig:rs-St_steep_V10}, presenting the radial grain size distribution for the steep disc with $\Vfrag=10$~m\,s$^{-1}$ and $\epsilon_0=1$, 3 and 5\% (Video~3, available online, shows its evolution) and in Fig.~\ref{Fig:Pmid_steep_e1_3_5_V10}, plotting the corresponding gas pressure profiles. After the same evolutionary time, grains have grown to larger sizes.

\begin{figure}
\centering
\resizebox{\hsize}{!}{
\includegraphics{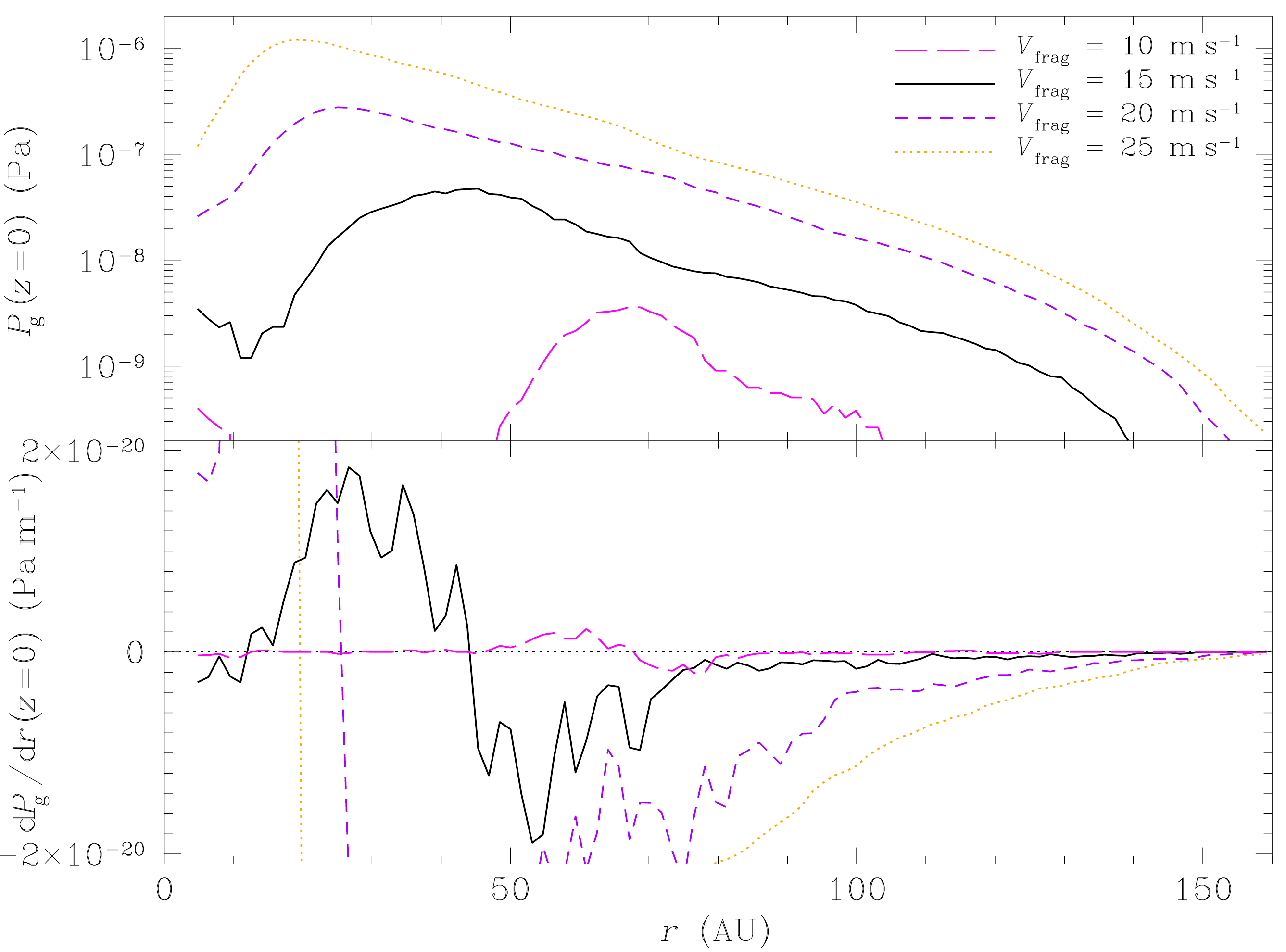}
}
\caption{Radial profiles of the gas pressure (top) and pressure gradient (bottom) in the midplane of the flat disc with $\epsilon_0=1$\% after 200,000~yr for $\Vfrag=10$, 15, 20 and 25~m\,s$^{-1}$.}
\label{Fig:Pmid_flat_V15_20_25}
\end{figure}

\begin{figure}
\centering
\resizebox{\hsize}{!}{
\includegraphics{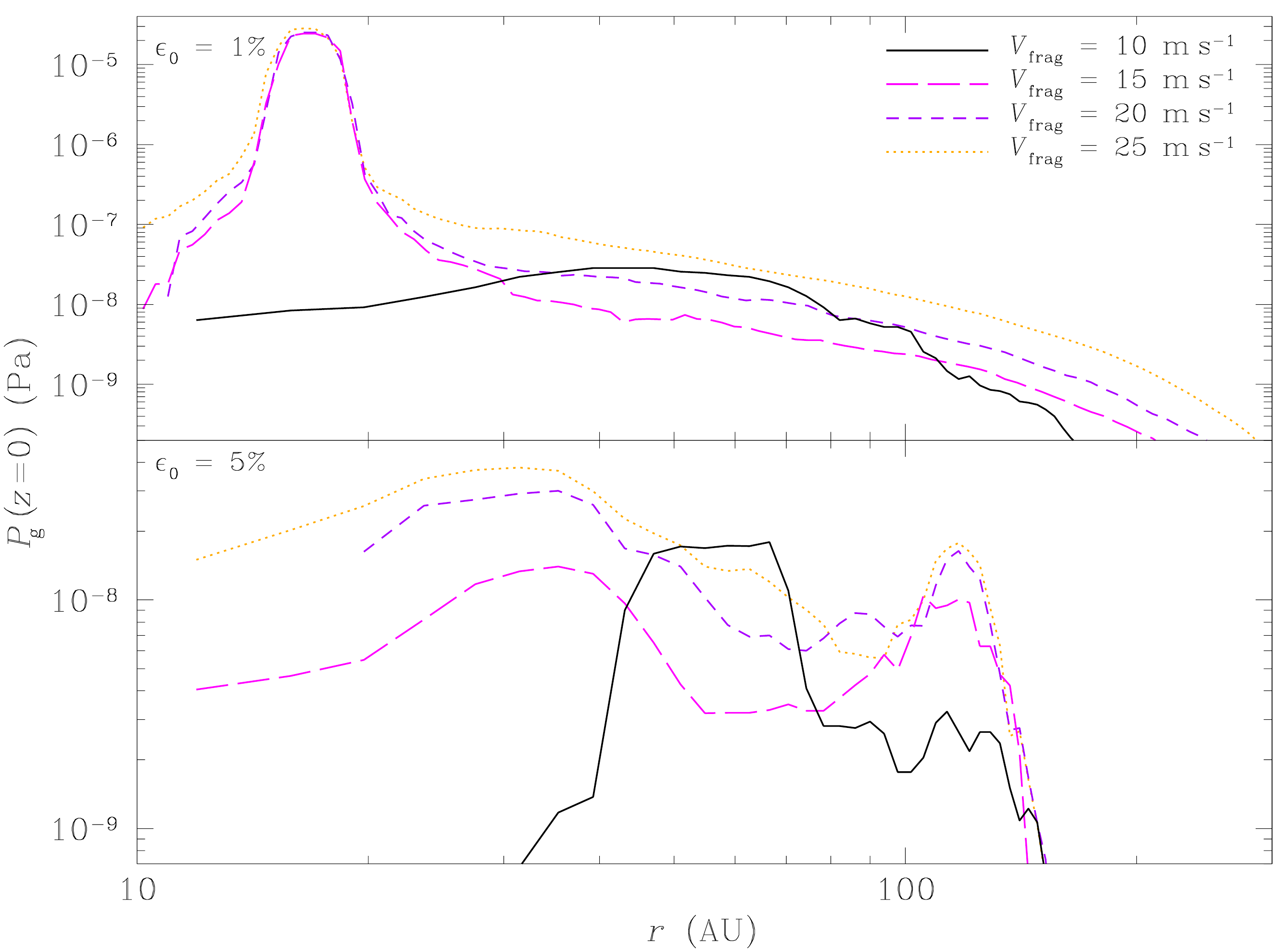}
}
\caption{Radial profiles of the gas pressure in the midplane of the steep disc after 400,000~yr for $\Vfrag=10$, 15, 20 and 25~m\,s$^{-1}$. Top: $\epsilon_0=1\%$. Bottom: $\epsilon_0=5\%$.}
\label{Fig:Pmid_steep_e1_5_V10_15_20_25}
\end{figure}

The radius exterior to which fragmentation is not effective varies as $r_\mathrm{frag}\propto\Vfrag^{-2/q}$ (equation~\ref{Eq:r_frag}). For larger fragmentation thresholds, grains can grow unhindered by fragmentation closer to the star, and therefore more rapidly due to the radial dependence of the growth time-scale. Those grains are the ones that eventually overcome the radial-drift and fragmentation barriers and pile up. The self-induced dust trap thus forms closer to the star and sooner. This is illustrated in Fig.~\ref{Fig:Pmid_flat_V15_20_25}, showing the locations of the gas pressure maxima for the flat disc with $\epsilon_0=1\%$ and values of $\Vfrag$ from 10 to 25~m\,s$^{-1}$. In the steep disc, the same trends are observed (see Fig.~\ref{Fig:Pmid_steep_e1_5_V10_15_20_25} and Video~4, available online) with some additional peculiarities. For $\epsilon_0=1\%$, when $V_{\rm frag}\ge15$~m\,s$^{-1}$, it is sufficiently large that grains manage to decouple very efficiently from the gas as they grow before reaching $r_\mathrm{frag}$. Their subsequent evolution is thus independent of $V_{\rm frag}$ and is the same as for growing grains when fragmentation is not taken into account, i.e. $V_{\rm frag}=+\infty$ \citep{Laibe2008,Laibe2014}. After a phase of rapid drift, they decouple from the gas and pile up in the inner disc. The self-induced dust trap thus forms at the same position for all values of $V_{\rm frag}\ge15$~m\,s$^{-1}$ (Fig.~\ref{Fig:Pmid_steep_e1_5_V10_15_20_25}, top). For $\epsilon_0=5\%$, grain growth is so fast that a first dust trap forms very early and at large distances, between 100 and 200~au. The gas compression at the location of the trap lowers the gas surface density, and therefore $s_\mathrm{frag}^+$, around it, which helps grains slightly interior to it to escape fragmentation. They drift inwards and grow, forming a second dust trap in the inner disc as discussed previously, at shorter radii for larger $\Vfrag$ (Fig.~\ref{Fig:Pmid_steep_e1_5_V10_15_20_25}, bottom). The formation of these two self-induced dust traps can only be seen in global simulations.

\section{Forming the trap}
\label{app:GasDyn}

\subsection{Equations of motion}

After a few stopping times, the radial velocity of the viscous gas phase of a dusty disc is given by
\begin{equation}
\begin{array}{r@{\ }c@{\ }l}
v_\mathrm{g} 
& = & -f_\mathrm{drag} \, \displaystyle\frac{1}{\Sigma_\mathrm{g} \Omega} \frac{\partial}{\partial r}\left(c_\mathrm{s}^2 \Sigma_\mathrm{g} \right) + \frac{\displaystyle\frac{\partial}{\partial r}\left( \Sigma_\mathrm{g} \nu r^3 \frac{\partial \Omega}{\partial r} \right)}{r \Sigma_\mathrm{g} \displaystyle \frac{\partial}{\partial r} \left( r^2 \Omega \right)}, \\
& \equiv & v_\mathrm{g,drag} + v_\mathrm{g,visc} .
\label{eq:vgr}
\end{array}
\end{equation}
$f_\mathrm{drag}$ is a dimensionless function of the Stokes number St and the dust-to-gas ratio $\epsilon$ given by
\begin{equation}
f_\mathrm{drag} \equiv \frac{\epsilon}{(1 + \epsilon)^2 \St^{-1} + \St} . \label{eq:fdrag}
\end{equation}
 The first term of the right-hand side of equation~\ref{eq:vgr} is the backreacting counterpart of the usual drift term obtained for a sub-Keplerian gaseous disc \citep{Nakagawa1986}. The minus sign indicates that dust backreaction makes the gas drift outwards by angular momentum conservation. The second term is the usual inwards viscous flow. In the zero-dust-mass limit $\epsilon = 0$, equation~\ref{eq:vgr} reduces to the seminal diffusion equation of a viscous disc \citep{LBP1974}. From equation~\ref{eq:vgr}, the conservation of mass reads
\begin{equation}
\frac{\partial \Sigma_\mathrm{g}}{\partial t} =  - \frac{1}{r} \frac{\partial }{\partial r}\left( \frac{\displaystyle\frac{\partial}{\partial r}\left( \Sigma_\mathrm{g} \nu r^3 \frac{\partial \Omega}{\partial r} \right)}{\displaystyle \frac{\partial}{\partial r} \left( r^2 \Omega \right)} \right) + \frac{1}{r} \frac{\partial}{\partial r} \left( f_\mathrm{drag}  \frac{r}{\Omega} \frac{\partial \left(c_\mathrm{s}^{2} \Sigma_\mathrm{g} \right)}{\partial r} \right). 
\label{eq:full_gas}
\end{equation}
The frequency $\Omega$ is the local Keplerian frequency $\Omega_\mathrm{K}$ corrected by perturbative terms according to
\begin{equation}
\Omega = \Omega_\mathrm{K} + \frac{1}{2 v_\mathrm{K} \Sigma_\mathrm{g}}\frac{\partial \left(c_\mathrm{s}^{2} \Sigma_\mathrm{g} \right)}{\partial r} .
\label{eq:def_omega}
\end{equation}
Looking at orders of magnitude, we denote $l$ the typical length over which surface density gradients develop in the disc, such that
\begin{align}
\frac{1}{v_\mathrm {K} \Sigma_\mathrm {g}}\frac{\partial \left(c_\mathrm{s}^{2} \Sigma_\mathrm{g} \right)}{\partial r}  & =   \mathcal{O}\left(\frac{H^{2}}{rl} \right) \Omega_\mathrm{K} , \label{eq:odg1} \\
\frac{1}{\Sigma_\mathrm{g} \Omega} \frac{\partial}{\partial r}\left(c_\mathrm{s}^2 \Sigma_\mathrm{g} \right) & = \mathcal{O}\left(\frac{H^{2}}{rl} \right) v_\mathrm{K} , \label{eq:odg2}\\
\frac{\displaystyle\frac{\partial}{\partial r}\left( \Sigma_\mathrm{g} \nu r^3 \frac{\partial \Omega}{\partial r} \right)}{r \Sigma_\mathrm{g} \displaystyle \frac{\partial}{\partial r} \left( r^2 \Omega \right)} & = \mathcal{O}\left(\alpha  \frac{H^{2}}{\min\left(r,l \right)^{2}} \right) v_\mathrm{K} \label{eq:odg3}.
\end{align}
When $l \gtrsim r$, equation~\ref{eq:odg3} is dominated by the global shear of the nearly-Keplerian rotation profile of the disc (the first term of the right-hand side of equation~\ref{eq:def_omega}). However, for $l \ll r $ the viscous velocity is dominated by a local shear induced by the correction to the orbital frequency (the second term in equation~\ref{eq:def_omega}, see \citealt{Kanagawa2015}). It should be noted that an additional viscous correction has been neglected in equation~\ref{eq:def_omega}. This approximation is valid as long as $l / r \ge \alpha $, a condition satisfied in practice. 

The relative importance of backreaction is measured via the parameter $x_\mathrm{br}$ defined as
\begin{equation}
x_\mathrm{br} \equiv \frac{\left| v_\mathrm{g,drag} \right| }{ \left| v_\mathrm{g,visc} \right| } \simeq \frac{1}{\alpha} \frac{(\min\left(l,r \right))^2}{lr} f_\mathrm{drag}.
\label{eq:def_xbr}
\end{equation}
 To estimate $x_{\rm br}$, one should distinguish two cases, depending whether a local pressure maximum has developed in the disc or not. This is similar to gap formation in dusty discs \citep{Dipierro2016}.

\subsection{Formation of the pressure maximum}

\begin{figure}
\centering
\resizebox{\hsize}{!}{
\includegraphics{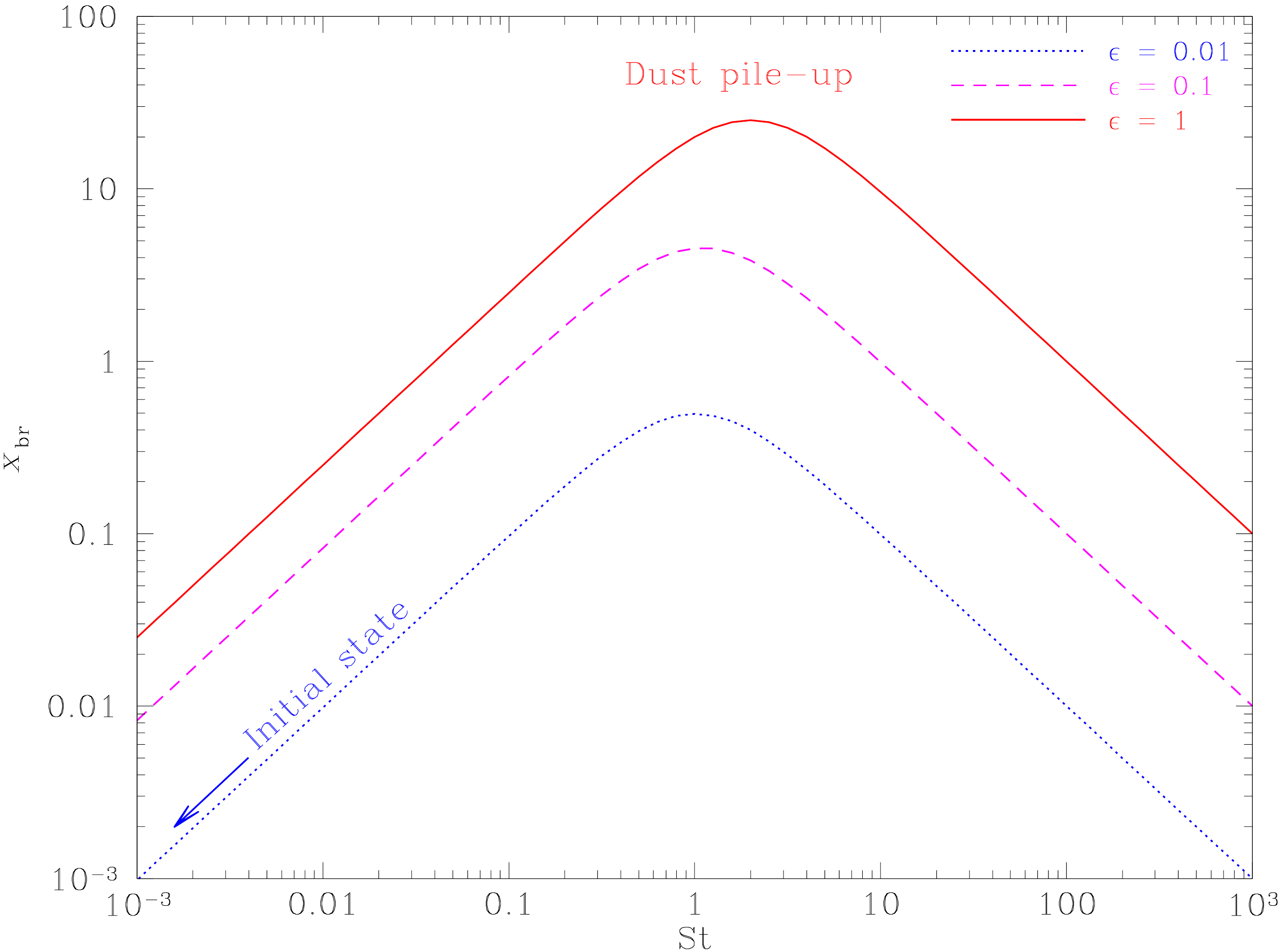}
}
\caption{Parameter $x_\mathrm{br}$, quantifying the importance of backreaction on the gas motion, as a function of $\St$, for $\epsilon=0.01$, 0.1 and 1, and $\alpha = 10^{-2}$.}
\label{Fig:x_br}
\end{figure}

We consider first a disc in which no pressure maximum has formed yet. This implies $l \simeq r$. Equation~\ref{eq:full_gas} reduces therefore to
\begin{equation}
\frac{\partial \Sigma_\mathrm{g}}{\partial t} =  \frac{3}{r} \frac{\partial }{\partial r}\left( \sqrt{r} \frac{\partial}{\partial r} \left( \sqrt{r}\nu\Sigma_\mathrm{g} \right) \right) + \frac{1}{r} \frac{\partial}{\partial r} \left( f_\mathrm{drag}  \frac{r}{\Omega_\mathrm{K}} \frac{\partial \left(c_\mathrm{s}^{2} \Sigma_\mathrm{g} \right)}{\partial r} \right) ,
\label{eq:full_gas_1}
\end{equation}
and an estimate of the parameter $x_{\rm br}$ is
\begin{equation}
x_\mathrm{br} \simeq \frac{f_\mathrm{drag}}{\alpha}.
\label{eq:x_br_1}
\end{equation}
Using equation~\ref{eq:fdrag}, the values of $x_{\rm br}$ as a function of $\St$ and $\epsilon$ for our value of $\alpha = 10^{-2}$ are shown in Fig.~\ref{Fig:x_br}.
Initially, $\epsilon \simeq 10^{-2}$ and $\St \ll 1$ in the entire disc: $f_\mathrm{drag}\sim\epsilon\,\St$ and $x_\mathrm{br}\ll1$. As expected, viscosity governs the gas evolution. Once dust has piled-up such that locally, $\epsilon\sim1$ and $\St\sim1$. $f_\mathrm{drag}$ is of order unity, implying a large $x_\mathrm{br}\sim1/\alpha$: $v_\mathrm{g,visc}$ becomes negligible, and backreaction takes over. This simple result is counter to most of the literature to date: gas dynamics can be dominated by dust, rather than by its viscosity. This is a local effect, since $f_\mathrm{drag}$ is a strongly peaked function around the dust concentration.
\begin{figure}
\centering
\resizebox{\hsize}{!}{
\includegraphics{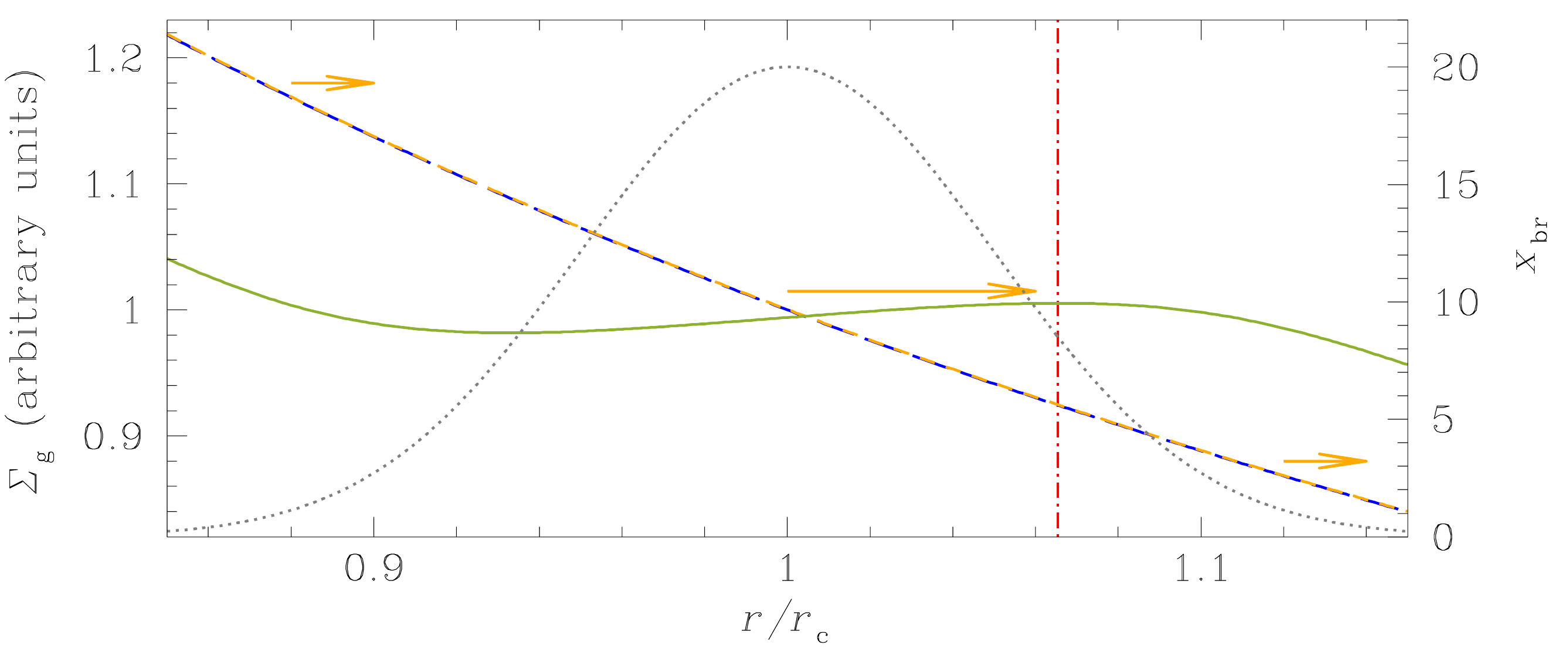}
}
\caption{Gas surface density profiles after $0.1$ viscous time at the dust concentration radius. The grey dotted line plots $x_\mathrm{br}$, representing the dust concentration (vertical scale on the right). With $x_\mathrm{br}^0=0.01$ and $x_\mathrm{br}^\mathrm{peak}=0$ (orange short-dashed line), the profile is almost superimposed with the purely viscous self-similar solution (blue long-dashed line). In the profile for  $x_\mathrm{br}^0=0.01$ and $x_\mathrm{br}^\mathrm{peak}=20$ (solid green line), a maximum forms close to the peak's inflection point with negative slope located at $r/r_\mathrm{c} = 1.065$ (vertical red dot-dashed line).}
\label{Fig:maxgas}
\end{figure}

To detail how backreaction modifies the gas structure at the location of a radial dust concentration, we integrate equation~\ref{eq:full_gas} numerically between $r/r_\mathrm{c} = 0.1$ and $r/r_\mathrm{c} = 100$, using a typical temperature profile of $T \propto r^{-0.5}$ and a peaked profile of $x_\mathrm{br}=x_\mathrm{br}^0+x_\mathrm{br}^\mathrm{peak}\mathrm{e}^{-\frac{(r-r_\mathrm{c})^2}{2\Delta r_\mathrm{c}^2}}$, with $x_\mathrm{br}^0 = 0.01$, $x_\mathrm{br}^\mathrm{peak} = 20$ and $\Delta r_\mathrm{c}/r_\mathrm{c} = 0.05$. We choose a conservative value of 20 for $x_\mathrm{br}^\mathrm{peak}$ and $\Delta r_\mathrm{c}$ is the typical width of the dust peak. We adopt for $\Sigma_\mathrm{g}$ the values of the self-similar solution at the boundaries \citep{LBP1974}, and the disc is evolved on a short time-scale to study the effect of the dust radial concentration on the gas structure (this formalism is not appropriate for long-term evolution since the dust surface density varies). Fig.~\ref{Fig:maxgas} shows the evolution of gas profiles obtained with $x_\mathrm{br}^\mathrm{peak}=0$ and $x_\mathrm{br}^\mathrm{peak}=20$. In the absence of a peaked dust distribution, $x_\mathrm{br}^\mathrm{peak}=0$ and the gas evolution is viscous and is almost identical to the self-similar solution. With a peaked dust distribution, here $x_\mathrm{br}^\mathrm{peak}=20$, mass is transferred from the inner edge of the pile-up, where a density minimum forms, to its outer edge, creating a density maximum. The disc is not dissipative enough to prevent the deformation of the gas profile, even with the large value of $\alpha\simeq10^{-2}$ used here. It may be noted that in the limiting case of a disc with very low dissipation ($\alpha<5\times10^{-3}$), the density maximum may induce the formation of a vortex by the Rossby wave instability \citep{Zhu2014}. For $q = 0.5$, we find that the gas profile becomes `bumpy' for $x_\mathrm{br}^\mathrm{peak}\gtrsim9$. Qualitatively, the typical radial scale of the pile-up is much shorter than the radial extension of the disc. Thus, an approximate equation of the gas evolution is
\begin{equation}
\frac{\partial \Sigma_\mathrm{g}}{\partial t}  \simeq \frac{1}{\Omega_\mathrm{K}} \frac{\partial}{\partial r} \left( c_{\rm s}^{2} \Sigma_\mathrm{g} \right) \frac{\partial f_{\rm drag}}{\partial r} .
\label{eq:approx}
\end{equation}
From equation~\ref{eq:approx}, the typical formation time of the gas density maximum is
\begin{equation}
\tau\sim\frac{r_\mathrm{c}\,\Delta r_\mathrm{c}}{H^2}\,t_\mathrm{K},
\label{eq:tau}
\end{equation}
where $H$ is the disc pressure scale height and $t_\mathrm{K}$ the Keplerian time-scale, and depends on the history of the dust evolution, according to the choice of disc and fragmentation parameters. For our conservative value of $\Delta r_\mathrm{c}/r_\mathrm{c}=0.05$ and a typical $H/r_\mathrm{c}\sim0.1$, $\tau\sim5\,t_\mathrm{K}$. Equation~\ref{eq:approx} shows that the gas density increases (resp. decreases) close to the inflection point of the function $f_\mathrm{drag}$ with a negative (resp. positive) slope.

Once a pressure maximum has formed in the gas, $l \ll r$. Local shear generates strong viscosity which acts to spread out the local density peak (a mechanism also discussed in \citealt{Yang2010} for Rayleigh-unstable discs) and prevent a too large concentrations of gas. The pressure maximum is stable and the gas flow is almost stationary when viscosity counterbalances the backreaction. This stable situation corresponds to $x_\mathrm{br} \simeq 1$. Near the pressure maximum, $v_\mathrm{g} \simeq 0$. The inwards flow of particles coming from the disc outer regions sustains an effective backreaction over long times at the trap location.

\section{Captions of online videos}

\noindent\textbf{Video~1:} Evolution of the radial grain size distribution in the flat disc with $\Vfrag=15$~m\,s$^{-1}$ and $\epsilon_0=1$\% up to 200\,000~yr with (top) or without (bottom) backreaction of dust on gas. The colour represents the Stokes number St (left) or the ratio $\Vrel/\Vfrag$ (right). Initially, grain growth is only efficient in the very outer disc, where grains have the lowest $\Vrel$, but is slow (its time-scale varies as $r^{3/2}$ \citep{Laibe2008}). With backreaction, the drift velocity of grains is slightly lower, delaying their drift to regions where growth is faster, they therefore grow more slowly than without backreaction. After $\sim70,000$~yr, their size distribution has caught up to the case without backreaction outside of $\sim100$~au. In the inner regions, the size distributions in both cases differ greatly because grains decouple from the gas or not when backreaction is included or not, respectively. With backreaction, most grains are piled-up in the self-induced dust trap and grow. Without backreaction, grains in the inner disc have high $\Vrel$ and fragment, replenishing the small grains reservoir. Grains thus stay well coupled to the gas and remain in the disc.
\\

\noindent\textbf{Video~2:} Same as Video~1 for the steep disc with $\Vfrag=15$~m\,s$^{-1}$ and $\epsilon_0=1$\% evolved up to 400\,000~yr.
\\

\noindent\textbf{Video~3:} Evolution of the radial grain size distribution in the steep disc with $\Vfrag=10$~m\,s$^{-1}$ and $\epsilon_0=1$, 3 and 5\% (from top to bottom) up to 400\,000~yr. The colour represents the Stokes number St (left) or the ratio $\Vrel/\Vfrag$ (right).
\\

\noindent\textbf{Video~4:} Evolution of the radial grain size distribution in the steep disc with $\epsilon_0=1\%$ (left) or 5\% (right) and $\Vfrag=10$, 15, 20 and 25~m\,s$^{-1}$ (from top to bottom) up to 400\,000~yr. The colour represents the Stokes number St.

\bsp	
\label{lastpage}
\end{document}


\label{firstpage}
\pagerange{\pageref{firstpage}--\pageref{lastpage}}
\maketitle


\begin{keywords}
errata, addenda -- hydrodynamics -- methods: numerical -- protoplanetary discs
\end{keywords}



After publication of the paper `Self-induced dust traps: overcoming planet formation barriers' in MNRAS 467, 1984 (2017), we discovered an inconsequential error in Appendix~C.

\begin{figure}
\centering
\resizebox{\hsize}{!}{
\includegraphics{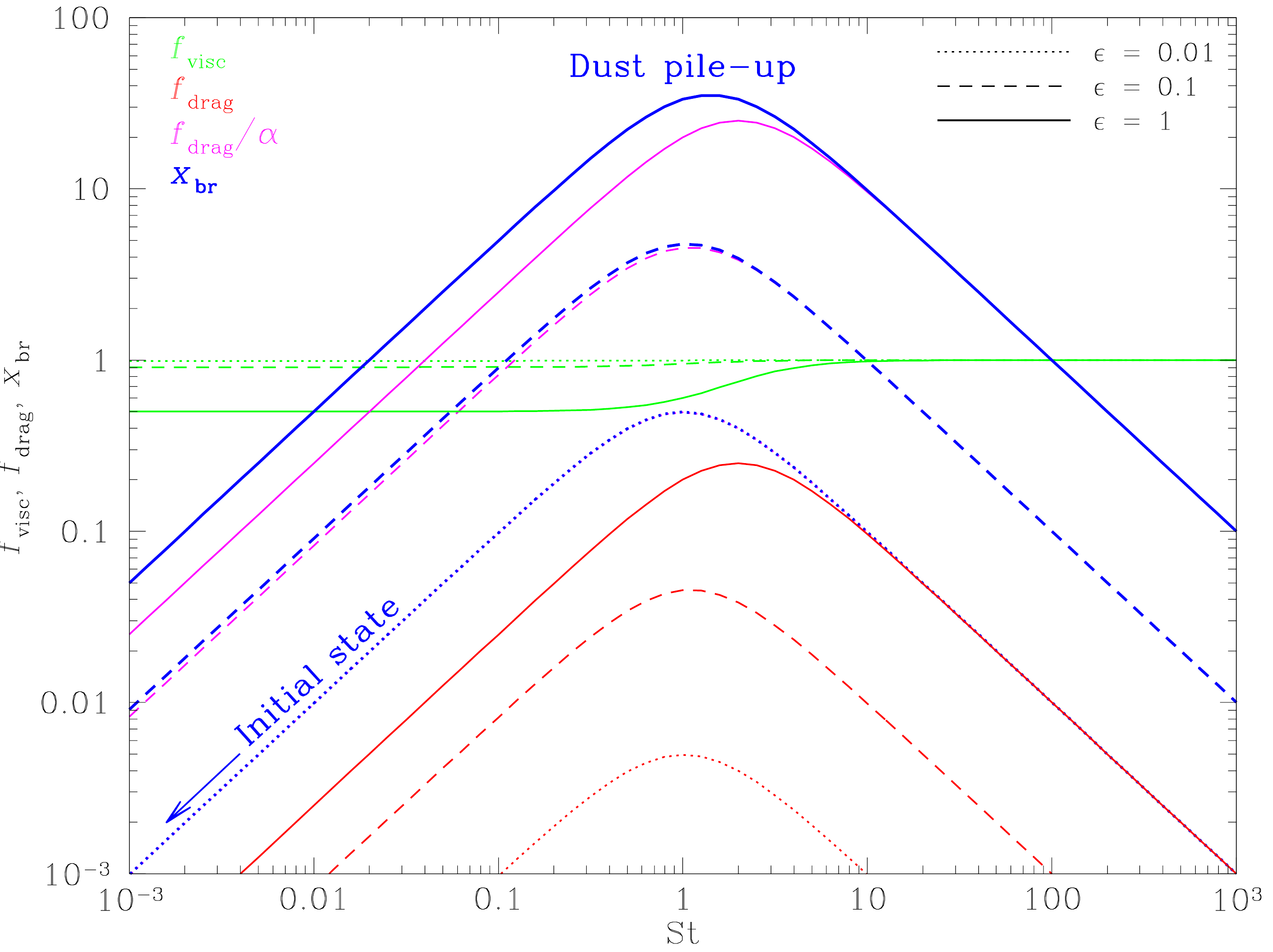}
}
\caption{Parameter $x_\mathrm{br}$, quantifying the importance of back-reaction on the gas motion, as well as $f_\mathrm{drag}$ and $f_\mathrm{visc}$, as a function of $\St$, for $\epsilon=0.01$, 0.1 and 1, and $\alpha = 10^{-2}$.}
\label{Fig:x_br}
\end{figure}

Steady-state expressions of the radial velocities for both the gas and dust phases of a dusty disc, taking into account the back-reaction of dust on gas, have been calculated by several authors, for different conditions \citep[e.g.][]{Nakagawa1986,Kretke2009,Dipierro2017,Kanagawa2017}. They are usually written as functions of the Stokes number for the gas-dust mixture $\St'$ and the dust-to-gas ratio $\epsilon=\rho_\mathrm{d}/\rho_\mathrm{g}$. In equation~(C1), we gave the steady-state expression of the radial velocity of the viscous gas phase of a dusty disc as a function of the more commonly used Stokes number defined for the dust only, $\St$, related to $\St'$ by $\St'=\St/(1+\epsilon)$ \citep{Price2015}:
\begin{equation}
\begin{array}{r@{\ }c@{\ }l}
v_\mathrm{g} 
& = & -f_\mathrm{drag} \, \displaystyle\frac{1}{\Sigma_\mathrm{g} \Omega} \frac{\partial}{\partial r}\left(c_\mathrm{s}^2 \Sigma_\mathrm{g} \right) + f_\mathrm{visc} \, \frac{\displaystyle\frac{\partial}{\partial r}\left( \Sigma_\mathrm{g} \nu r^3 \frac{\partial \Omega}{\partial r} \right)}{r \Sigma_\mathrm{g} \displaystyle \frac{\partial}{\partial r} \left( r^2 \Omega \right)}, \\
& \equiv & v_\mathrm{g,drag} + v_\mathrm{g,visc} ,
\label{eq:vgr}
\end{array}
\end{equation}
where
\begin{equation}
f_\mathrm{drag} = \frac{\epsilon}{(1 + \epsilon)^2 \St^{-1} + \St} . \label{eq:fdrag}
\end{equation}
When replacing $\St'$ by its expression as a function of $\St$, we mistakenly wrote that $f_\mathrm{visc}=1$. The correct expression is
\begin{equation}
f_\mathrm{visc} = \frac{(1 + \epsilon)\,\St^{-1} + \St}{(1 + \epsilon)^2 \St^{-1} + \St} . \label{eq:fvisc}
\end{equation}
As a consequence, equation~(C10) should now read
\begin{equation}
x_\mathrm{br} \equiv \frac{\left| v_\mathrm{g,drag} \right| }{ \left| v_\mathrm{g,visc} \right| } \simeq \frac{1}{\alpha} \frac{f_\mathrm{drag}}{f_\mathrm{visc}},
\label{eq:def_xbr}
\end{equation}
and Fig.~C1 should be replaced with the current Fig.~\ref{Fig:x_br}. It shows that both $f_\mathrm{visc}$ and $x_\mathrm{br}$ are little affected for small values of $\epsilon$. For $\epsilon\sim1$, the effect of back-reaction on the gas phase is up to twice as large, strengthening the conclusions of the initial study.

%



\bibliographystyle{mnras}
\bibliography{SelfTraps_Erratum}


\bsp	
\label{lastpage}